\documentclass[journal]{IEEEtran}
\usepackage{etoolbox}
\usepackage{caption} 
\makeatletter
\patchcmd{\@makecaption}
{\scshape}
{}
{}
{}
\makeatletter
\patchcmd{\@makecaption}
{\\}
{:\ }
{}
{}
\makeatother

\ifCLASSINFOpdf
\else
\fi
\usepackage{graphicx}

\usepackage[implicit=false]{hyperref}
\usepackage{amsmath}
\usepackage{amsfonts}
\usepackage{amssymb}
\newcommand{\I}{\iota}
\usepackage{xcolor}
\usepackage{marginnote}
\usepackage{authblk}
\newcommand{\rev}[1]{{\color{black}{#1}}}
\newcommand{\revv}[1]{{\color{black}{#1}}}
\newcommand{\revvv}[1]{{\color{blue}{#1}}}
\newcommand{\tamirrev}[1]{{\color{black}{#1}}}
\newcommand{\noamrev}[1]{{\color{black}{#1}}}
\newcommand{\noam}[1]{{\color{black}{#1}}}
\newcommand{\nrev}[1]{{\color{black}{#1}}}
\newcommand{\nrevv}[1]{{\color{black}{#1}}}
\newcommand{\secrev}[1]{{\color{black}{#1}}}
\usepackage[section]{placeins}
\newcommand\blfootnote[1]{%
	\begingroup
	\renewcommand\thefootnote{}\footnote{#1}%
	\addtocounter{footnote}{-1}%
	\endgroup
}
\usepackage{algorithmic}
\usepackage{algorithm}
\begin{document}
%\title{Improving the Accuracy and Efficiency of Multi-Reference Alignment Using Synchronization}
%\title{Consolidating Synchronization With Expectation Maximization in the Multi-Reference Alignment Problem}
\title{An accelerated expectation-maximization \secrev{algorithm} for multi-reference alignment}
% \title{Bare Demo of IEEEtran.cls\\ for IEEE Journals}/

\author{Noam~Janco and 
        Tamir~Bendory% <-this % stops a space
}
\maketitle

\begin{abstract}
\rev{
The multi-reference alignment (MRA) problem entails
estimating an image from {multiple} noisy and rotated copies
of itself. 
If the noise level is low, one can reconstruct the image
by estimating the missing rotations, aligning the images, and averaging
out the noise. 
While accurate rotation estimation is impossible if the noise level is high, 
the rotations can still be {approximated, and thus \tamirrev{can} provide indispensable information}.
{In particular, learning the approximation error can be harnessed for efficient image estimation.}
{In this paper, we }propose a new {computational} framework, called Synch-EM, that \tamirrev{consists} of angular synchronization followed by expectation-maximization (EM). 
The synchronization step results in a concentrated distribution of
rotations; this distribution is learned and then incorporated
into the EM as a Bayesian prior. The learned distribution also {dramatically} reduces
the search space, and thus the computational load of the EM iterations. We show by extensive numerical
experiments that the proposed framework can \tamirrev{significantly} accelerate EM
for MRA\tamirrev{ in high noise levels, occasionally by a few orders of magnitude,} {without degrading the reconstruction quality.}
}
%In this work, we propose a framework called Synch-EM, that comprises of coarse synchronization preceding to the EM  which results in concentrated distribution of rotations. This distribution can be learned and incorporated into the EM by  introducing a prior on the distribution of rotations after synchronization, reducing the search space and as an initial model. 
%\begin{itemize}
%	\item Prior on the distribution of rotations
%	\item Reducing the search space
%	\item Initial model
%\end{itemize}
%The suggested framework can significantly reduce the runtime without increasing the reconstruction error.

%In this work, we propose a framework to consolidate synchronization techniques with the existing reconstruction algorithms in a variety of SNRs. By learning the statistical properties of the observations after synchronization, we suggest modified algorithms with prior on the distribution of rotations that can significantly reduce the runtime without increasing the reconstruction error.
\end{abstract}


% \fi
\IEEEpeerreviewmaketitle

\section{Introduction}

\IEEEPARstart{M}{ulti}-reference alignment (MRA) is an abstract mathematical model inspired by the problem of \rev{constituting} a 3-D molecular structure using cryo-electron microscopy (cryo-EM). The MRA problem entails estimating an image from multiple noisy and rotated copies of itself. The \tamirrev{computational} and statistical properties of the MRA problem have been analyzed thoroughly in recent years, {see} \cite{BispectrumInversion,
	bandeira2020optimal,perry2019sample,b2013multireference,AperiodicMRA,2dMRA,boumal2017heterogeneous, bandeira2020non, bandeira2017estimation, romanov2021multi, abas2022generalized, bendory2021sparse, ghosh2021multi, bendory2022dihedral, liu2021decompose}. 
%\subsection{2-D Multi-reference alignment}
\blfootnote{\revv{N. Janco and T. Bendory are with the school of Electrical
	Engineering, Tel Aviv University, Israel, (e-mail: noamjanco@gmail.com; bendory@tauex.tau.ac.il). The research is partially
	supported by} \nrev{the NSF-BSF grant award 2019752, the BSF grant no. 2020159, the ISF grant no. 1924/21, and the Zimin Institute for Engineering Solutions Advancing Better Lives.}}

This paper focuses on the 2-D MRA model. Let \(I\) be an image of size \(L \times L\) that we wish to estimate. 
%The pixels in the image are indexed by a pair of integers \((x,y)\) with \(-(L-1)/2\leq x,y\leq (L-1)/2\).
Let \(R_\theta\) be a rotation operator which rotates an image counter-clockwise by an angle \(\theta\).
The observations are i.i.d. random samples \rev{drawn} from the model:
\begin{equation}
\label{MRA_model}
Y = R_{\theta} I + \varepsilon,
\end{equation}
where the rotations are \tamirrev{drawn from  a distribution \(\theta\sim\rho\),} %distributed %according to some distribution
and \tamirrev{$\varepsilon$ is a noise matrix
whose entries are i.i.d.\ Gaussian variables with zero mean and variance $\sigma^2$.} 
% \(\varepsilon=(\varepsilon_{ij})\in \mathbb{R}^{L\times L}\) is a noise matrix with zero mean and variance \(\sigma^2\). 
The variables \(\theta\) and \(\varepsilon\) are independent.
Inspired by cryo-EM, we assume that \(\theta\) is distributed uniformly over \(\left[0,2\pi \right)\) \cite{bendory2020single}.
%In cryo-EM datasets it is common to assume \(\xi \sim \mathcal{U}\left[0,2\pi \right)\) since all angles of projections are equally likely. 
%Furthermore, in practice, the range of possible rotations is discretized by L possible values in the range \(\left[0,2\pi \right)\).
%The notation of general distribution of rotations is introduced so that the model will hold for synchronized data as well.
The task is to estimate the unknown image~\revv{\(I\)} from \revv{a} set of noisy images \rev{\(Y_1,...,Y_N\)}, while \revv{the rotations} \(\theta_1,...,\theta_N\) are unknown. \tamirrev{We note that model~\eqref{MRA_model} suffers from an  unavoidable ambiguity of rotation, and therefore, naturally, a solution is defined up to a rotation.}
%\revvv{The solution is defined up to a rotation due to the symmetry of the problem.}

Following \cite{2dMRA, zhao2016fast}, \rev{we assume that the image can be represented in a steerable basis with finitely many coefficients, namely\revv{,}
%\begin{equation}
%\label{fb_expansion}
%I(r,\theta)=\sum_{k,q}{a_{k,q}u^{k,q}(r,\theta)},
%\end{equation}
\noam{
\begin{equation}
\label{fb_expansion}
I(r,\theta)=\sum_{k=-L}^L\sum_{q=1}^{p_k}{a_{k,q}u^{k,q}(r,\theta)},
\end{equation}}
%where \(\alpha_{k,q} \) are the coefficients of the expansion, and \(k,q\) are the angular and radial frequencies, respectively. The steerable basis function in polar coordinates takes the form \(u^{k,q}(r,\theta)=f^{k,q}(r)e^{\iota k\theta}\), where \(f^{k,q}(r)\) is a real valued radial function and \(\iota=\sqrt{-1}\), 
where \(u^{k,q}(r,\theta)=f^{k,q}(r)e^{\iota k\theta}\) is \revv{a} steerable basis function in polar coordinates, \(f^{k,q}(r)\) is a real valued radial function, \(\iota=\sqrt{-1}\), \noam{and \(p_k\) denotes the number of components 
%satisfying \(R_{k,q}\leq \pi L/2\)
according to \nrev{a} sampling criterion \nrev{described in} Appendix~\ref{FBCoeffs}}. The \nrev{expansion coefficients} are denoted by~\(a_{k,q} \), where \(k,q\) are the angular and radial {indices}, respectively.
}
%we expand the images with respect to a steerable basis in order to effectively handle image rotations. In polar coordinates, steerable basis functions take the form:
%\begin{equation}
%u^{k,q}(r,\theta)=f^{k,q}(r)e^{\iota k\theta},
%\end{equation}
%where \(f^{k,q}(r)\) is a real valued radial function. We assume that the image can be expanded in \(u^{k,q}\):
%\begin{equation}
%\label{fb_expansion}
%I(r,\theta)=\sum_{k,q}{\alpha_{k,q}u^{k,q}(r,\theta)},
%\end{equation}
%where \(\alpha_{k,q} \) are the coefficients of the expansion and \(k,q\) are the angular and radial frequencies, respectively.
\rev{Since \(u\) is a steerable basis, rotating the image by} an angle \(\phi\) corresponds to an addition of a linear phase in the coefficients:
%\begin{equation}
%\label{fb_rotation}
%I(r,\theta-\phi)=\sum_{k,q}{a_{k,q}u^{k,q}(r,\theta-\phi)}
%=\sum_{k,q}{a_{k,q}e^{-\I k\phi}u^{k,q}(r,\theta)}.
%\end{equation}
\noam{
\begin{equation}
\label{fb_rotation}
\begin{aligned}
I(r,\theta-\phi)&=\sum_{k=-L}^L\sum_{q=1}^{p_k}{a_{k,q}u^{k,q}(r,\theta-\phi)}
\\
&=\sum_{k=-L}^L\sum_{q=1}^{p_k}{a_{k,q}e^{-\I k\phi}u^{k,q}(r,\theta)}.
\end{aligned}
\end{equation}
}
\tamirrev{As} we consider real images, the coefficients satisfy a symmetry, \noam{\(a_{k,q}=\overline{a_{-k,q}}\)}, \revv{and therefore} coefficients with~\(k\geq0\) suffice to represent the images.
%\rev{In particular, we use the Fourier-Bessel basis. The definition of the Fourier-Bessel and the computation of the coefficients of the observed images are described in Appendix \ref{FBCoeffs}.
%Accordingly, we expand all observable images \(Y_1,...,Y_N\) in Fourier-Bessel basis. In addition,} to reduce the dimensionality of the representation and to denoise the image, steerable PCA (sPCA) is applied after expanding the images in the Fourier-Bessel basis \cite{zhao2016fast}. Similarly to standard PCA, sPCA results in a new data driven basis to represent the images, while preserving the steerability property.
%Notably, the orthonormality of the Fourier-Bessel and sPCA bases implies that the noise statistics are preserved.
In particular, we use the steerable PCA (sPCA) basis \cite{zhao2016fast}, \tamirrev{which results in a data driven, steerable,} 
basis to represent the images. \nrev{In this case, the relation (2) approximately holds; see Appendix~\ref{FBCoeffs}.} 
%Thus, the observed coefficients are i.i.d. samples from the model:
%\begin{equation}
%\label{MRA_Model_FB}
%v_{k,q} = a_{k,q}e^{-\I k\theta}+\varepsilon^c_{k,q},
%\end{equation}
%where \(a_{k,q}\) are the sPCA coefficients of the unknown image~\(x\), \(\theta \sim\ \mathcal{U}\left[0,2\pi\right)\), and \(\varepsilon^c\) denotes a complex Gaussian vector, satisfying \(\mathbb{E}\varepsilon^c=0\) and \(\mathbb{E}\varepsilon^c(\varepsilon^c)^*=\sigma^2 I\), where \(I\) is the identity matrix. 
%The image formation model is described in detail in Section \ref{Image Formation Section}. 

\rev{Based on our image formation model\tamirrev{~\eqref{fb_expansion}}, we reformulate the 2-D MRA problem as follows.}
Each 2-D MRA observation can be described as:
\begin{equation}
\label{MRA_Model_FB}
v_i = a \circ e^{-\I \bar{k}\theta_i} + \varepsilon^c_i, \quad i=1,\ldots,N,
\end{equation}
\rev{where \(\circ\) denotes entry-wise product,}
\(a\) is the sPCA coefficients vector of the unknown image~\(I\), \(v_i\) is the sPCA coefficients vector of the $i$-th observation, {\(\theta_i\in\left[0,2\pi\right)\) is the unknown \tamirrev{corresponding }rotation,}
%\(\theta \sim\ \mathcal{U}\left[0,2\pi\right)\), 
and \(\varepsilon_i^c\) denotes a complex Gaussian vector
%satisfying \(\mathbb{E}\varepsilon_i^c=0\) and \(\mathbb{E}\varepsilon_i^c(\varepsilon_i^c)^*=\sigma^2 I_{M\times M}\), where \(I_{M\times M}\) is the identity matrix. 
\revv{with i.i.d. entries with zero mean and variance \(\sigma^2\).}
We treat the sPCA coefficients \(\{v_{k,q}\}\) as a vector \(v\in\mathbb{C}^M\) such that \(v[j] = v_{\bar{k}[j],\bar{q}[j]}, \quad j=1,...,M\), where \nrev{\((\bar{k}, \bar{q}) \in \mathbb{Z}^M\)} specify the angular and radial {indices} \noamrev{at each entry}, and \(M\) is the number of sPCA coefficients used in the representation.
\rev{Therefore, given} a set of noisy sPCA coefficient vectors \(v_1,...,v_N,\) \rev{corresponding to} the noisy images \(Y_1,...,Y_N\), we wish to estimate the sPCA coefficients vector \(a\) of the unknown image \(I\)\tamirrev{, up to a global rotation.}  Once the sPCA coefficients vector is estimated, the image can be reconstructed using the sPCA basis.

Previous \rev{works on the MRA model} focused \revv{on two} signal to noise (SNR) regimes: high SNR (small \(\sigma\)) and extremely low SNR (\(\sigma\to \infty\)). 
\rev{{In the sequel}, we define \(\text{SNR} = \frac{\|I\|_\text{F}^2}{L^2\sigma^2}\).}
In \rev{the} high SNR \rev{regime}, 
one can estimate the unknown rotations using the \rev{angular} synchronization framework \cite{singer2009angular, Boumal_2016, Perry_2018, bandeira2020non}. \rev{Then,} the observations can be aligned (i.e., undo the rotations) and averaged; see Section \ref{HighSNR} for details.
However, in very low SNR\revvv{,} accurate rotation estimation is impossible, and thus methods 
%such as expectation-maximization (EM) \cite{ExpectationMaximization} 
that circumvent rotation estimation were developed and analyzed. 
%These methods are popular and have been proven to be effective for cryo-EM and MRA. Both of these methods are formulated by different non-convex optimization problems.
In particular, this paper {studies} \rev{expectation-maximization (EM)} \cite{ExpectationMaximization},
%This paper focused on EM, 
the most successful methodology for \rev{MRA and} cryo-EM \revv{\cite{scheres2012relion, sigworth1998maximum}}, which is \rev{outlined} in Section \ref{LowSNR}.

\begin{figure}[H]
	\includegraphics[width=\linewidth]{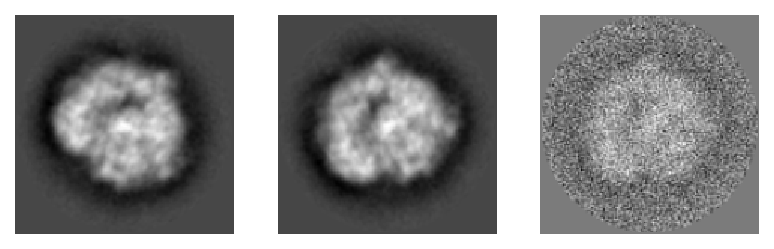}
	\caption{From left to right: The original clean image of a projection \nrev{image} of the E. coli 70S ribosome volume \cite{Shaikh2008}, \rev{a} rotated clean image, and \rev{a} noisy image with SNR=1/10.}
	\label{MRA_Observations_Figure}
\end{figure}

In practice, the SNR of cryo-EM data (the main motivation of this work) is not high enough for accurate estimation of rotations, but also not extremely low. \rev{A} typical SNR ranges between 1/100 to 1/10; this regime is exemplified in Figure~\ref{MRA_Observations_Figure}.
%Some of the most common scenarios of cryo-EM are in the intermediate regime, where the SNR is not high enough to apply high SNR methods, but not infinitely low. 
In this low (but not extremely low) SNR, synchronization provides indispensable information about the rotations that was neglected \tamirrev{by} previous works. 
In particular, while we cannot estimate the rotations accurately, we can estimate the \emph{distribution of rotations} after synchronization.  

In this paper\nrev{,} we present the Synch-EM framework (\tamirrev{see} Section \ref{Synch-EM}), in which synchronization \rev{is} incorporated into EM in order to accelerate the reconstruction process. 
Synchronization helps accelerating EM in three different \tamirrev{ways}. 
First, it allows us to find a good initialization point, which is important since the \revv{MRA} problem~\eqref{MRA_model} is non-convex; similar ideas are {also} \rev{described in the cryo-EM literature}~\cite{greenberg2017common,bandeira2020non}. 
Second, we can establish a meaningful prior information on the distribution of rotations after synchronization. The prior is comprised of a similarity function between the estimated distribution and the expected distribution of rotations, \revv{which} is learned;
%over different data
see Section \ref{PriorOnDistributionSection} for details. 
Third, if we have confidence that the distribution of rotations is concentrated (depending on the SNR), then we \nrev{can} reduce the search space (occasionally, dramatically), and thus reduce the computational burden; see Section \ref{ReducedSearchSpaceSection}.
To mitigate statistical dependencies (discussed in Section \ref{SynchProblems}) and to process massive datasets, we \rev{also} suggest a new synchronization framework\revvv{,} called Synchronize and Match (see Section \ref{SynchAndMatchSection}).

\section{Image Recovery in High and Low SNR Regimes}
\label{ImageRecovery}
This paper focuses on image recovery in \rev{the} low, but not extremely low, SNR regime. Before presenting the proposed framework, we introduce the classical methods in high and low SNRs and their shortcomings.   
%Then, we discuss the mid-range SNR regime, which is the main focus of this paper and the typical SNR in cryo-EM datasets.
\subsection{High SNR} \label{HighSNR}
In high SNR, one can \rev{reliably} estimate the missing rotations \(\theta_1,...,\theta_N\), \rev{undo} the rotations, %\(R_{\hat{\theta}_i}^{-1}Y_i\)
and recover the image by averaging. 
%We present two common methods to estimate rotations.
%Both of these methods rely on estimating the relative rotations between pairs of images:
\rev{G}iven any two observations \(Y_i, Y_j\) drawn from~(\ref{MRA_model}), \tamirrev{their} relative rotation \(\theta_{i,j}\) can be estimated by
\begin{equation}
\hat{\theta}_{i,j} =\arg\min_{\theta\in [0,2\pi)}{\|Y_i-R_{-\theta} Y_j\|^2}.
\end{equation}
This can be written in terms of the sPCA coefficients of the images \(v^i_{k,q}\), \(v^j_{k,q}\):
\begin{equation}
\label{relative_angle_2d_dist}
\hat{\theta}_{i,j}
= \arg{
	\min_{\theta\in [0,2\pi)}{\sum_{k,q}{|v^i_{k,q} - e^{\I k\theta}v^j_{k,q}|^2}}
}.
\end{equation}
\rev{We now present two methods to estimate rotations.}
%Many techniques to estimate the rotations in the high SNR regime were devised. The observations can then be aligned and averaged in order to get an estimation of the underlying signal.
\subsubsection{Template matching}
A simple, albeit not robust to noise, method to estimate rotation\rev{s} is by template matching.
In this method, all observations are aligned \rev{with respect} to a single reference image. 
%The alignment of each observation to the common reference is done by estimating the relative rotation between them, as will be described later. 
The computational complexity of template matching is \(O(NLM)\), where \(N\) is the number of observations, \(L\) is the number of \revv{possible angles (discretization of the unit circle \(\left[0,2\pi\right)\))}, and \(M\) is the number of sPCA coefficients. The template matching estimator works well only in high SNR scenarios and tends to be biased; see for example \cite{Shatsky2009}, which show\rev{s} that a reference image (e.g.\rev{,} a portrait of Einstein) can be extracted from pure random noise \revv{images}.
% due to strong dependency on the reference image.

\subsubsection{\rev{Angular} Synchronization}
\label{SynchronizationSection}
The \tamirrev{angular} synchronization framework provides a more robust computational paradigm as it considers the ratios between all pairs of images. 
%In synchronization, the rotation of each image is estimated according to the relative rotations between all pairs. The synchronization comprises of two steps:
%Estimating rotations from their relative ratios: the set of angles \(\theta_1,..,\theta_N \) is estimated from their 
%	offset \(\theta_i-\theta_j \quad mod \quad 2\pi\).
%estimated offsets \(\hat{\theta}_{i,j}\).
If the estimated rotations between all pairs are arranged in a matrix \revv{\(H\in\mathbb{C}^{N\times N}\)} such that \(H_{i,j}=e^{\iota\hat{\theta}_{i,j}}\), the angular synchronization problem can be associated with the maximum likelihood estimation problem:
%	\begin{equation}
%	\max_{\theta_1,...,\theta_N \in [0,2\pi)}
%	{\sum_{i,j=1}^N{e^{-\I \theta_i}H_{ij}e^{-\I \theta_j}}
%	}
%	\end{equation}
%	which can be rewritten as:
\begin{equation}
\label{cost_synchronization}
\max_{z\in \mathbb{C}_1^N}{z^*Hz},
\end{equation}
where \( \mathbb{C}_1^N := \{z\in\mathbb{C}^N:|z_1|=...=|z_N|=1\}\). This is a smooth, non-convex optimization problem on the manifold of product of circles.
%	This problem can be solved in different ways, such as convex relaxation followed by power iterations, or by a projected power method which iterates until convergence according to:
To solve (\ref{cost_synchronization}), we follow \cite{Boumal_2016} and apply the projected power method (PPM){, which is an iterative algorithms whose {$(t+1)$ iteration reads}}:
\begin{equation}
	z^{(t+1)}=\text{phase}(Hz^{(t)}),
\end{equation}
where \(\text{phase}(z)_i=z_i/|z_i|\).
The estimated rotations are then:
\begin{equation}
\hat{\theta}_i = \arg{z_i} \quad \rev{i=1,...,N},
\end{equation}
\rev{where \(\arg()\) is the argument of a complex number.}
The computational complexity of this method is \rev{governed} by the computation of the  relative rotations of all pairs. Since we have \(N(N-1)/2\) distinct pairs, the overall complexity is \(O(N^2LM)\), an order of \(N\) higher than template matching.
%Since computing the relative rotations between each pair of images scales as \(O(N^2)\), the computational load becomes huge for large \(N\).
The dependency on \(N^2\) implies that the computational burden becomes prohibitive for large \(N\). 
{Therefore, a} modification of the synchronization algorithm must be made in order to process big datasets in \rev{a} reasonable time.

Once the rotations are estimated (either by template matching or PPM), each image can be easily {aligned} using the steerability property, \(\hat{v}^{i}_{k,q}=e^{\I k\hat{\theta}_i}v^{i}_{k,q}\). Thus, the sPCA coefficients of the image \tamirrev{are estimated by }averaging the aligned coefficients:
\begin{equation}
\label{synch_initialization}
\hat{a}_{k,q} = \frac{1}{N}\sum_{i=1}^{N}{e^{\I k\hat{\theta}_i}v^{i}_{k,q}}.
\end{equation} \revv{From the estimated sPCA coefficients, one can reconstruct the image using the sPCA basis.}

\subsection{Expectation-maximization}
\label{LowSNR}
When the SNR is sufficiently \rev{low}, accurate rotation estimation is impossible. Therefore, in this regime, techniques that estimate the signal directly
%, while circumventing rotation estimation, 
are used. A popular approach is to maximize the posterior distribution after marginalizing over the rotations:
%In this paper, we wish to maximize the marginalized posterior distribution:
%\begin{equation}
%x^*,\rho^* = \arg \max_{x,\rho} p(\boldsymbol{y} | x,\rho),
%\end{equation}
\begin{equation}
a^*,\rho^* = \arg \max_{a,\rho} p(a,\rho|\boldsymbol{v}),
\end{equation}
where \(a, \rho\) are, respectively, the unknown image coefficients and \revv{the} distribution of rotations, and 
\(\boldsymbol{v}=(v_1,...,v_N)\in \mathbb{C}^{M\times N}\) is the observation matrix \tamirrev{whose} \revv{$i$-th} column corresponds to the sPCA coefficients of \revv{the $i$-th} observation~\eqref{MRA_Model_FB}.
% with \(M\) denoting the number of coefficients.
%we refer to this estimator as the maximum marginalized likelihood estimator (MMLE). 
%The MMLE of the signal and the distribution of rotations in the MRA problem is given by:
%The MMLE for the signal \(x\) and the distribution of rotations \(\rho\), given the observation matrix \(\boldsymbol{y}=(y_1,...,y_N)\) according to the observation model in Eq. (\ref{MRA_model}), is the maximizer of \(p(\boldsymbol{y}|x,\rho)\).
\rev{Recall that (now {without marginalizing out the rotations})}, the posterior distribution is proportional to the likelihood times the prior:
\begin{equation}
p(a,\rho|\boldsymbol{v},\boldsymbol{\theta}) \propto p(\boldsymbol{v},\boldsymbol{\theta}|a,\rho)p(a)p(\rho),
\end{equation}
where \(p(a)\) is the prior \rev{on} the image coefficients \(a\)\revvv{,} and \(p(\rho)\) is the prior \rev{on} the distribution of rotations \(\rho\), assuming \(a\) and \(\rho\) are independent. 
In particular, the likelihood function of 2-D MRA\nrev{,} in terms of sPCA coefficients~\eqref{MRA_Model_FB}\nrev{,} is given by:
	\begin{equation}
		p(\boldsymbol{v},\boldsymbol{\theta}|a,\rho)
		=
		\prod_{j=1}^{N}\rho[\theta_j]
		\frac{1}{\pi^M\sigma ^{2M}}
		e^{-\frac{1}{\sigma^2}\|a\circ e^{-\I \rev{\bar{k}}\theta_j}-v_j\|_2^2}\rev{.}
\end{equation}
Importantly, up to now, the prior on the distribution of rotations \(p(\rho)\) was \rev{overlooked by} previous works in the MRA and cryo-EM literature \cite{scheres2012relion, BispectrumInversion, AperiodicMRA, 2dMRA, bendory2020super, sigworth1998maximum}.
%when the posterior probability was considered. %The prior on the distribution \(\rho\), depends on the prior on the signal x. 
%Thus, they are not independent. However, they can be approximated as independent.

%The likelihood function for 2-D MRA (\ref{MRA_model}) is given by:
%\begin{equation}
%p(\boldsymbol{y},\boldsymbol{\theta}|x,\rho)
%=
%\prod_{j=1}^{N}\rho[\theta_j]
%\frac{1}{(2\pi\sigma ^2)^{L^2/2}}
%e^{-\frac{1}{2\sigma^2}\|R_{\theta_j}x-y_j\|_2^2}.
%\end{equation}
%The likelihood function of 2-D MRA in terms of sPCA coefficients (\ref{MRA_Model_FB}) is given by:
%\begin{equation}
%p(\boldsymbol{v},\boldsymbol{\theta}|a,\rho)
%=
%\prod_{j=1}^{N}\rho[\theta_j]
%\frac{1}{\pi^M\sigma ^{2M}}
%e^{-\frac{1}{\sigma^2}\|a\circ e^{-\I \rev{\bar{k}}\theta_j}-v_j\|_2^2}\rev{.}
%\end{equation} %\marginnote{\(\bar{k}\) is described in (4)}
%where \(a\) and \(v_j\) are the sPCA coefficients vector of the unknown image and the \(j\)-th observation,  respectively, \(M\) is the number of sPCA coefficients in the signal expansion and \(k \in \mathbb{Z}^M\) specifies the angular frequency at each entry of the coefficient vectors.
%The F-B coefficients are flattened into a one dimensional vector in \(\mathbb{C}^M\), and \(k \in \mathbb{Z}^M\) specifies the angular frequency at each entry of the coefficient vectors.

The \rev{EM} algorithm aims to \rev{iteratively} maximize the marginalized posterior distribution as follows \rev{\cite{dempster1977maximum}}.
First, 
%the range of possible rotations must be discretized by L possible values in the range \(\left[0,2\pi \right)\) 
we fix a range of \(L\) possible rotations \([0,\frac{2\pi}{L},...,\frac{2\pi(L-1)}{L}]\),
according to \revv{a} required resolution. Then, given a current estimate of the image coefficients \(a_t\) and distribution \(\rho_t\), consider the expected value of the log posterior function, with respect to the conditional distribution of \(\theta\) given \(\boldsymbol{v}\), \(a_t\) and \(\rho_t\):
%\begin{equation}
%\label{expectation_step}
%Q(x,\rho|x_t,\rho_t)=E_{\theta|\boldsymbol{y},x_t,\rho_t}\{\log{p(x,\rho|\boldsymbol{y},\boldsymbol{\theta})}\}
%\end{equation}
%\begin{equation}
%\label{expectation_step}
%\begin{aligned}
%Q(x,\rho|x_t,\rho_t)&=E_{\theta|\boldsymbol{y},x_t,\rho_t}\{\log{p(x,\rho|\boldsymbol{y},\boldsymbol{\theta})}\}
%\\
%&=
%\sum_{j=1}^{N}{\sum_{l=0}^{L-1}{w_t^{l,j}\{\log\rho[l]-\frac{1}{2\sigma^2}\|R_{\frac{2\pi l}{L}}x-y_j\|^2\}}}
%\\
%&+\log p(x) +\log p(\rho)
%\end{aligned},
%\end{equation}
\begin{equation}
\label{expectation_step}
\begin{aligned}
Q(a,\rho|a_t,\rho_t)&=\rev{\mathbb{E}}_{\theta|\boldsymbol{v},a_t,\rho_t}\{\log{p(a,\rho|\boldsymbol{v},\boldsymbol{\theta})}\}
\\
&=
\sum_{j=1}^{N}{\sum_{l=0}^{L-1}{w_t^{l,j}\{\log\rho[l]-\frac{1}{\sigma^2}\|a \circ e^{-\I\frac{2\pi  l}{L}\bar{k}}-v_j\|^2\}}}
\\
&+\log p(a) +\log p(\rho).
\end{aligned}
\end{equation}
The weight \(w_t^{l,j}\) is the probability that the rotation \(\theta_j\) of observation \(j\) is equal to \(\frac{2\pi l}{L}\), given \(\boldsymbol{v}\) and assuming \(a=a_t,\rho=\rho_t\), and is \rev{proportional to}:
%\begin{equation}
%w_t^{l,j}\propto e^{-\frac{1}{2\sigma^2}\|R_{\frac{2\pi l}{L}}x_t-y_j\|^2}\rho_t[l]
%\end{equation}
%or equivalently:
\begin{equation}
\label{weights_2d_em}
w_t^{l,j}\propto e^{-\frac{1}{\sigma^2}\|a_t\circ e^{-\I\frac{2\pi  l}{L}\bar{k}}-v_j\|^2}\rho_t[l].
\end{equation}
Then, \(a\) and \(\rho\) are updated by:
\begin{equation}
\label{maximization_step}
(a_{t+1},\rho_{t+1}) = \arg{\max_{a,\rho}{Q(a,\rho|a_t,\rho_t)}}.
\end{equation}

The \tamirrev{EM} algorithm iterates between the expectation (\ref{expectation_step}) and \revv{the} maximization (\ref{maximization_step}) steps until convergence.
We assume throughout that the image coefficients were drawn from a complex Gaussian distribution with zero mean and covariance \(\Gamma_a\) such that \(p(a)={1/\left(\pi^M\det{(\Gamma_a)}\right)}\exp(-a^H\Gamma_a^{-1}a)\), \revv{where~\((\cdot)^H\) denotes the conjugate transpose operator.} \tamirrev{In this case,} the maximizer over the coefficients reads:
\begin{equation}
\label{coefficients_update}
a_{t+1} = 
\left(NI + \sigma^2\Gamma_a^{-1}\right)^{-1}\sum_{j=1}^{N}{\sum_{l=0}^{L-1}
	{
		w_{t}^{l,j}e^{\I\frac{2\pi l}{L}\bar{k}}\circ v_j
	}
}.
\end{equation}
%When there is no prior on the image (i.e., uninformative prior), the term \(\sigma^2\Gamma_a^{-1}\) vanishes.
When there is no prior on the distribution of rotations, the update rule of the distribution reads:
\begin{equation}
\label{distribution_update}
\rho_{t+1}[l] = \frac{W_t[l]}
{\sum_{l=0}^{L-1}{W_t[l]}},
\end{equation}
where \(W_t[l]=\sum_{j=1}^N{w_t^{l,j}}\). The derivations of (\ref{coefficients_update}) and (\ref{distribution_update}) are given in Appendix \ref{Mstep}.
The computational complexity of \revv{EM} is \(O(TNLM)\), where \(T\) is the number of EM iterations, which tends to increase as the SNR decreases.

\begin{figure}[t]
	% generated with plot_cov_after_synch.m
	\includegraphics[width=\linewidth]{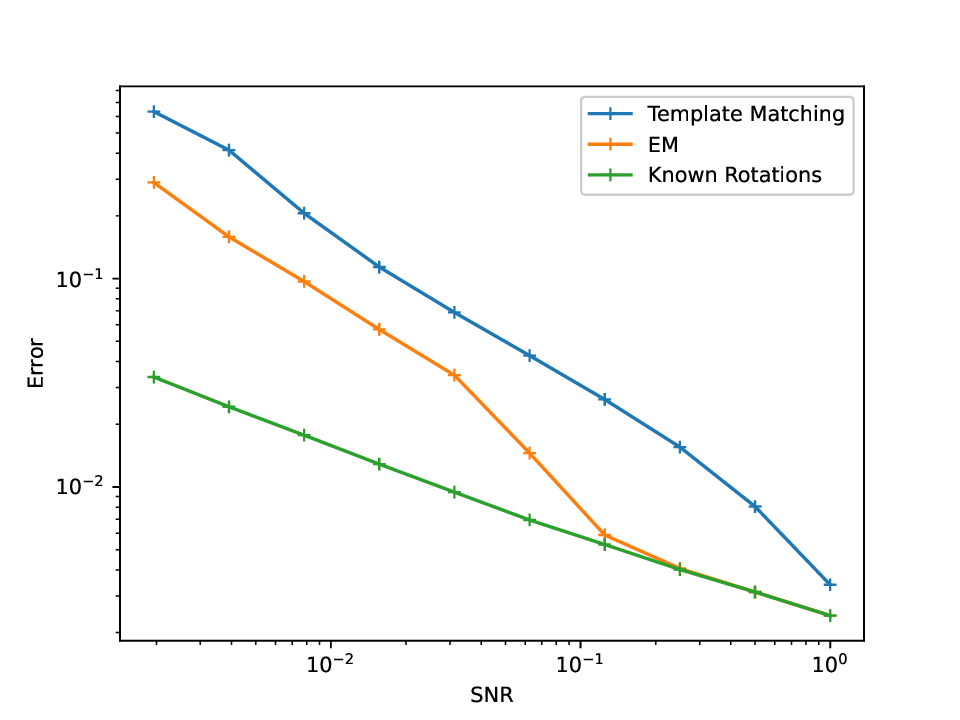}
	\caption{The relative error of template matching, \rev{EM} and \revv{the case of known rotations}, in different SNRs, with \(N=5000\) samples of 2-D MRA with projection images of the E. coli 70S ribosome.}
	\label{RelErr_comparison}
\end{figure}

Figure~\ref{RelErr_comparison} shows the \revv{relative} error as a function of the SNR of EM, template matching, and MRA with known rotations (which is equivalent to averaging \rev{i.i.d.} Gaussian variables). 
{To account for the rotation symmetry, the} relative error between an estimated image \(\hat{I}\) and the true image \(I\) is defined as:
\begin{equation}
\text{relative error}=\min_{\theta\in [0,2\pi)}\frac{\|R_{\theta}\hat{I}-I\|_\text{F}}{\|I\|_\text{F}}.
\end{equation} 
In this experiment, we use projection images of the E. coli 70S ribosome volume taken from \cite{Shaikh2008}. Each image is of size \(129\times 129\) pixels. For each SNR \rev{value}, we ran 10 \rev{trials} with \rev{different projection images} and \(N=5000\) observations.
The rotations were drawn \rev{uniformly} from a set of 360 possible angles \(\frac{2\pi}{360}[1,...,360]\), rather than a continuous range over~\(\left[0,2\pi\right)\). 
This allows us to avoid an error floor in high SNR when setting \(L=360\).
We computed the sPCA coefficients according to \cite{zhao2016fast}, in which the number of sPCA coefficients  is \rev{automatically chosen as a function of the SNR.}
We ran the EM without a prior on the image coefficients.
As can be seen,
in the high SNR regime\revv{,} accurate rotation estimation is possible: the performance\rev{s} of template matching and EM \rev{are} \rev{competitive} to \revv{the case of known rotations}. 
However, \tamirrev{low SNR hampers accurate rotation estimation.}
%when the SNR drops, \rev{accurate} rotation estimation is impossible.
 In this regime, EM that directly estimates the image outperforms template matching by a large margin.
 
\section{Accelarated EM using synchronization}
\label{Synch-EM}
The Synch-EM framework consists of three main ingredients. 
The first step involves synchronizing the \rev{observations}. However, synchronization algorithms induce correlations between the signal and the noise terms of the aligned images, \revv{which in turn leads to} poor performance of the EM (which assumes \rev{model} (\ref{MRA_model})). To mitigate this effect, we \rev{propose a new synchronization paradigm, called} Synchronize and Match (Section \ref{SynchAndMatchSection}). Secondly, the expected distribution of rotations after synchronization is learned (Section \ref{DistributionLearningMethodologySection}). Finally, an EM algorithm, \rev{which takes the learned prior into account,} is applied to the synchronized observations. The learned distribution is also used to narrow the search space (Section \ref{ReducedSearchSpaceSection}), leading to {a} \revv{notable} acceleration.
The framework is outlined in Algorithm \ref{Synch-EM algorithm}.

\begin{algorithm}
	\caption{Synch-EM Algorithm}
	\label{Synch-EM algorithm}
	\begin{algorithmic}[1]
		\renewcommand{\algorithmicrequire}{\textbf{Input:}}
		\renewcommand{\algorithmicensure}{\textbf{Output:}}
		\REQUIRE 
		\begin{itemize}
			 \item[]
%			\item \makebox[1.5cm]{\(y_1,...,y_N\)\hfill}  A set of observations  drawn from (\ref{MRA_model}) 
%			\item \makebox[1.5cm]{\(L\)\hfill} Number of possible rotations
%			\item \makebox[1.5cm]{\(\overline{\rho}\)\hfill} Expected rotation distribution after synchronization  learned by Algorithm \ref{Algorithm 1}
%			\item \makebox[1.5cm]{\(\gamma\)\hfill} Prior regularization term 
%			\item \makebox[1.5cm]{\(BW\)\hfill} Search space bandwidth 
%			\item \makebox[1.5cm]{\(\Gamma_a\)\hfill} The covariance of the image coefficients  used as a prior
%			\item \makebox[1.5cm]{\(tol\)\hfill}Tolerance for stopping criteria 
%			\item \makebox[1.5cm]{\(T\)\hfill} Maximum number of iterations
%			\item \(Y_1,...,Y_N\) -  A set of observations  drawn from (\ref{MRA_model}) 
		\nrev{\item \(v_1,...,v_N\) -  A set of observations  drawn from (\ref{MRA_Model_FB}) }
			\item \(L\) - Number of rotations
			\item \(\overline{\rho}\) - Expected rotation distribution after synchronization  learned by Algorithm \ref{Algorithm 1}
			\item \(BW\) - Search space bandwidth%, chosen according to Algorithm~\ref{Algorithm 1} \marginnote{Currently I did not define how to choose it, it's a parameter chosen "by eye", can be chosen according to percentiles of the learned distribution}
			\item \(\gamma\) - Prior \noamrev{weight} term 

			\item \(\Gamma_a\) - The \revv{covariance of the \noamrev{image} prior}
			\item \(tol\) - Tolerance for stopping criteria 
			\item \(T\) - Maximum number of iterations
		\end{itemize}
		\ENSURE  \(\hat{I}\) - image estimate
%		\\ \textit{Initialization} :
%		\STATE \revv{Compute the set of sPCA coefficients per image \noamrev{\(\{\hat{v}_i\}_{i=1}^N\)}}
		\STATE Align the \noam{coefficients} according to Algorithm \ref{SynchAndMatch}, resulting in a new set of coefficients \noamrev{\(\{{\tilde{v}}_i\}_{i=1}^N\)}
		\STATE Initialize \noamrev{\(a_0=\frac{1}{N}\sum_{i=1}^N{\tilde{v}_i}\)}, \(\rho_0=\overline{\rho}\)
%		\\ \textit{LOOP Process}
		%TODO: SAY THAT RHO0 IS BETWEEN 0,BW-1
		\FOR {$t = 0 $ to $T-1$}
		\STATE Compute $
		w_t^{l,j}= \exp{\left(-\frac{1}{\sigma^2}\|a_t \circ e^{-\I\frac{2\pi  l}{L}\bar{k}}-\tilde{v}_j\|^2\right)}\rho_t[l]
		$ \noam{\(\forall l\in[-BW/2,BW/2]\)}, \(\forall j\in[1,...,N]\)
		\STATE Normalize $w_t^{l,j} \leftarrow \frac{w_t^{l,j}}{\noam{\sum_{l=-BW/2}^{BW/2}}{w_t^{l,j}}}$
		\STATE Compute \(W_t[l]=\sum_{j=1}^N{w_t^{l,j}}\)
		\STATE Update the \revv{coefficients} \\
%		$
%		a_{t+1} = 
%		\left(NI + \sigma^2\Gamma_a^{-1}\right)^{-1}\sum_{j=1}^{N}{\noam{\sum_{l=-\frac{BW}{2}}^{\frac{BW}{2}}}
%			{
%				w_{t}^{l,j}e^{\I\frac{2\pi l}{L}\bar{k}}\circ \tilde{v}_j
%			}
%		}
%		$
		\begin{align*}
		a_{t+1} =& \\
		&\left(NI + \sigma^2\Gamma_a^{-1}\right)^{-1}\sum_{j=1}^{N}{\noam{\sum_{l=-BW/2}^{BW/2}}
			{
				w_{t}^{l,j}e^{\I\frac{2\pi l}{L}\bar{k}}\circ \tilde{v}_j
			}
		}	
		\end{align*}
		\STATE \label{dist_update}Update the distribution $ \rho_{t+1}[l]= \frac{W_t[l]+\gamma\bar{\rho}[l]}
		{\noam{\sum_{l=-BW/2}^{BW/2}}{(W_t[l]+\gamma\bar{\rho}[l])}}$
		\IF {$\min_{l\in[0,L-1]}\|a_{t+1}-e^{-\frac{2\pi \I \bar{k}l}{L}}\circ a_{t}\|^2 < tol$}
		\STATE break
		\ENDIF
		\ENDFOR
		\STATE Reconstruct the estimated image \(\hat{I}\) from the estimated \revv{sPCA} coefficients \(a\) and the \revv{sPCA} basis.
%		\RETURN $\hat{x}$
	\end{algorithmic}
\end{algorithm}

%\subsection{Pros}
%\begin{itemize}
%	\item Initialization - after the images are synchronized, an initial estimate of the signal can be made by averaging over the synchronized images. The estimation can be used as an initialization to the following EM algorithm.
%	\item Concentrated distribution - previous work \cite{AperiodicMRA} proved that the relative error performance are superior in the case of a-periodic distribution compared to uniform distribution.
%	\item Prior on the distribution - the distribution of rotations after synchronization can be learned on similar data. Incorporating a prior on the distribution can add information, which might lead to improved performance.
%	\item Reduced search space - the concentrated distribution after synchronization allows to reduce the interval in which SO(2) is sampled. This reduction can lead to significant speed-up.
%\end{itemize}
\subsection{Synchronize and Match}
\label{SynchAndMatchSection}
In the first part of this section, we describe how applying synchronization \revv{to} the data affects the statistical model of the problem. Simulations show that the effect is \revv{substantial} and leads to poor performance if \revv{disregarded}.
In the second part, we introduce the Synchronize and Match algorithm that was designed to mitigate statistical dependencies induced by synchronization. In addition, it can be applied efficiently to large datasets.

\subsubsection{Statistical dependencies induced by synchronization}
\label{SynchProblems}
%\subsection{Statistical model after synchronization}
A synchronization algorithm receives a set of observations \rev{\(Y_1,...,Y_n\)} and outputs estimates of their corresponding rotations \(\hat\theta_1,...,\hat\theta_N\) (up to \revv{a} global rotation). Then, the observations are aligned:
\begin{equation}
\tilde{\rev{Y}}_i = R_{-\hat{\theta_i}}(R_{\theta_i}\rev{I} + \varepsilon_i)
%R_{\theta_i-\hat{\theta_i}}x + R_{-\hat{\theta_i}}\varepsilon_i,
=
R_{\Delta \theta_i}\rev{I}+\tilde{\varepsilon}_i,
\end{equation}
where \(\Delta \theta_i =\theta_i-\hat{\theta}_i\) and \(\tilde{\varepsilon}_i=R_{-\hat{\theta_i}}\varepsilon_i\)
are the \revv{synchronization error} and the noise term after alignment, respectively. \revv{The distribution of the synchronization error is concentrated} (later, in Section \ref{DistributionLearningMethodologySection}, we explain how this distribution can be learned). %\rev{Note that since synchronization} can estimate the rotations up to a global rotation, the distribution is centered around \rev{an arbitrary} angle \(\theta_c\).

%The problem can be reformulated as:
%\begin{equation}
%\label{MRA_model_after_synchronization}
%\tilde{y}_i = R_{\Delta \theta_i}x+\tilde{\varepsilon}_i.
%\end{equation}

Crucially, since \(\hat{\theta}_i\) is a function of the unknown rotations, the unknown signal and the noise realizations,
%\(\theta_i=f_i(\theta_1,...,\theta_N,x,\varepsilon_1,...,\varepsilon_N)\),
the noise term~\(\tilde{\varepsilon}_i\) is not necessarily distributed as \(\mathcal{N}(0,\sigma^2I)\), and is not necessarily uncorrelated with the signal. 
%For example, consider the case of 1-D MRA observations in low SNR, where \(\|x\|<\|\varepsilon_i\|\), and consider a synchronization method that for each observation \(y_i\) estimates \(\theta_i=\arg\max_l y_i[l]\), so that the signal \(y_i\) is rotated such that the maximum entry is the 0 entry. It is clear that \(E[\tilde{\varepsilon}_i[0]]> 0\).
%The bias of the noise after synchronization, \(E[R_{-\hat{\theta}_i}\varepsilon_i]\) had been examined for the 1-D case using a simulation, in which PPM synchronization was used with a signal \(x\sim\ \mathcal{N}(0,I)\). Simulations have shown that the mean of the noise vector is approximately \(\mu [l] = \sin \left(\frac{2\pi l}{L}+\phi \right)\), where \(\phi \sim\ U(0,2\pi)\), as can be seen in Figure \ref{induced_bias}.
%\begin{figure}[H]
%	\centering
%	\includegraphics[width=70mm]{ppm_noise_average}
%	\caption{Different realization of \(E[R_{-\hat{\theta}_i}\varepsilon_i]\), with \(L=21\),\(\sigma=3\),\(N=1000\).}
%	\label{induced_bias}
%\end{figure}
%In the following experiment, we use the Pearson correlation test to demonstrate that indeed synchronization induces correlation between the signal and the synchronized noise. 
To demonstrate that synchronization induces correlation between the noise and the shifted signals, we use the notion of Pearson correlation coefficients. 
Given a set of \rev{pairs of} samples \(\{(a_1,b_1),...,(a_N,b_N)\}\)\rev{,} the Pearson correlation coefficient between \(a\) and \(b\) is computed \revv{by}
\begin{equation}
r_{a b}=
\frac{\sum_{i=1}^N{(a_i-\bar{a})(b_i-\bar{b})}}{\sqrt{\sum_{i=1}^N{(a_i-\bar{a})^2}\sum_{i=1}^N{(b_i-\bar{b})^2}}},
\end{equation}
where \(\bar{a}\) and \(\bar{b}\) are, respectively, the sample mean of \(a\) and \(b\). Under the assumption that \(a\) and \(b\) are drawn \revv{independently} from normal distributions, the variable \(t=r\sqrt{\frac{N-2}{1-r^2}}\) follows a Student's t-distribution with \(N-2\) degrees of freedom. This also holds approximately \revv{for other distributions with} large enough sample size. 
This allows us to \rev{check whether~\(a\) and~\(b\) are independent variables with} {some} prescribed significance level (say, \(\alpha=0.05\)). 
%Using the inverse cumulative distribution function, we can compute \(r_{critical}\),  a threshold for \(|r|\), that corresponds to the significance level \(\alpha\).
%By setting a p-value threshold of \(\alpha=0.05\), we can compute the corresponding \(r_{critical}\) using the inverse cumulative distribution function. Values of \(|r|\) that exceed \(r_{critical}\) are unlikely for independent variables with a significance level \(\alpha\). 

\begin{figure}
	% generated with plot_cov_after_synch.m
	\includegraphics[width=\linewidth]{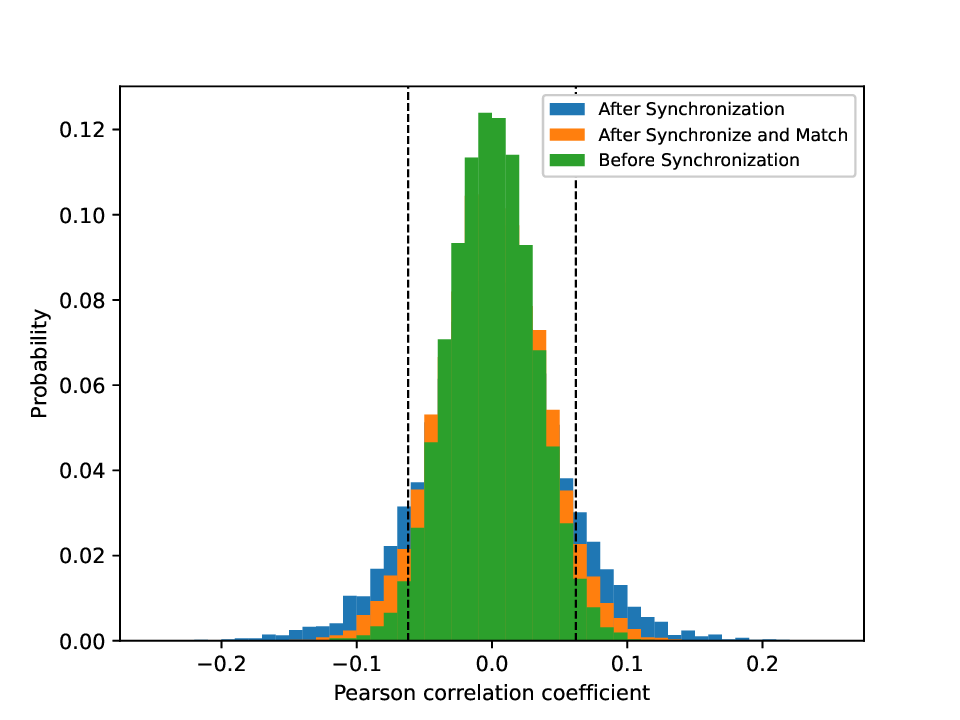}
	\caption{{C}orrelation can be detected using the Pearson correlation test. The \rev{panel} shows normalized histogram\rev{s} of the Pearson correlation coefficients \revv{between} the shifted signal and noise, before and after synchronization. The dashed vertical lines correspond to \rev{a significance level of \(\alpha=0.05\) according to the Student's t-distribution}. The experiment confirms that while the shifted signal and noise before synchronization are uncorrelated, synchronization introduces a significant amount of correlation between them. Synchronize and Match \nrev{(Algorithm~\ref{Synch-EM algorithm})} alleviates this problem.}
	\label{PearsonTest}
\end{figure}

For this demonstration, and to avoid the effects of {change of basis}, we use the 1-D discrete version of~(\ref{MRA_model}). In this case, the sought signal is a 1-D discrete vector \(x\in\mathbb{R}^L\), and \(R_\theta\) is a circular shift operator where the shifts are uniformly distributed over \([0,...,L-1]\).
%Similarly to the 2-D MRA, it is common to examine the problem with
%uniformly distributed shifts.
Here we use a signal \revv{of length \(L=21\),} \rev{whose entries were drawn independently from a normal distribution with zero mean and unit variance}\nrev{. We set} \(\sigma=2\) and \(N=1000\).
For each MRA trial, we computed the Pearson coefficient between each \noamrev{entry of the} shifted signal and \revv{the} noise before synchronization: \(r_{R_sx[i],\varepsilon[j]}\), \(\forall i,j\in[0,L-1]\). We repeated the experiment for 20 trials. As expected, we measured a fraction of 0.048 of Pearson coefficients whose p-values are smaller than 0.05. This verifies that the shifted signal and the noise entries, before synchronization, are indeed uncorrelated. Then, we repeated the same experiment \revv{with} the shifted signal and the noise after \rev{angular synchronization, using PPM}. As can be seen in Figure~\ref{PearsonTest}, more entries have large Peasron coefficients, indicating that they are unlikely to be uncorrelated. We measured a fraction of 0.205 of Pearson coefficients whose p-values are smaller than 0.05. \tamirrev{Therefore,} we conclude that the shifted signal and noise are correlated after synchronization. 

\revv{To remedy the correlation problem}, this paper suggests a new synchronization method, introduced in the following section, called Synchronize and Match. Conducting the same experiment with this method alleviates the induced correlation\rev{: we measured} a fraction of 0.102  Pearson coefficients whose p-values are smaller than 0.05.

\begin{figure}
	% generated with mra_1d_parallel.py
	\includegraphics[width=\linewidth]{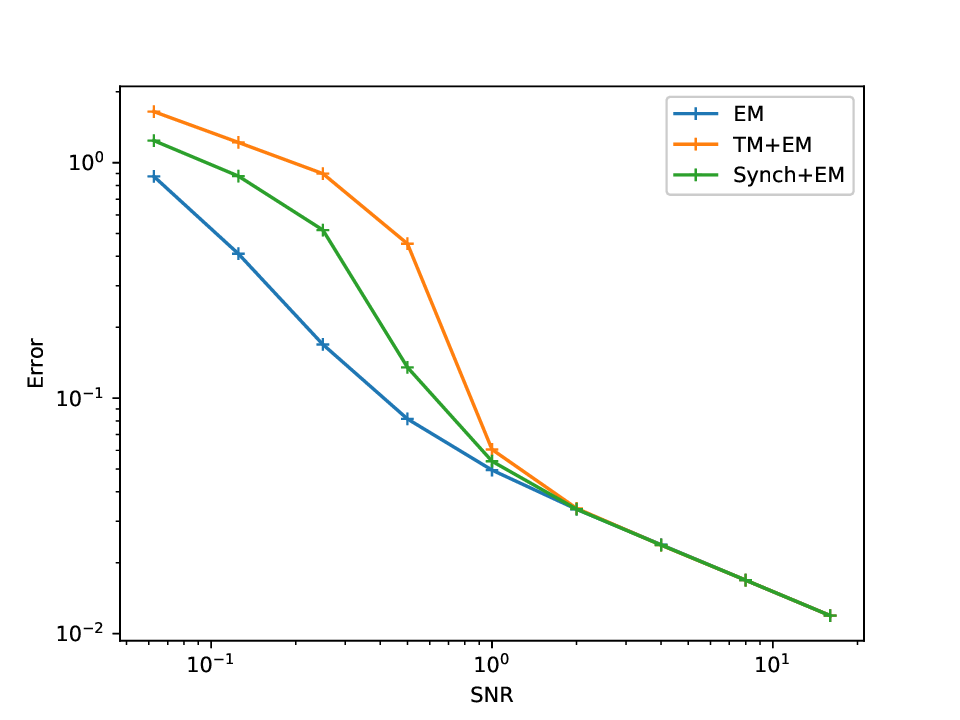}
	\caption{Degraded performance of EM after alignment, \rev{due to correlations induced by synchronization. In blue, standard EM; in orange, template matching followed by EM (TM+EM); and in green, angular synchronization using PPM} followed by EM (Synch+EM).}
	\label{degraded_performance}
\end{figure}

Figure \ref{degraded_performance} shows the relative error as a function of SNR of EM, template matching followed by EM, and \rev{angular synchronization using PPM} followed by EM.
In this \tamirrev{1-D MRA} experiment, \tamirrev{we used  a signal of length \(L=21\) whose entries were drawn independently from a normal distribution with zero mean and unit variance. For  each SNR, we ran 10 trials with \(N=500\).}
Notice how the correlation induced by synchronization degrades the performance of EM (which assumes the model~(\ref{MRA_model})).
%Therefore, to run an effective EM algorithm, 
%alignment methods that mitigate the correlation between the noise and the shifted signal must be considered.
%\subsection{Synchronization over large dataset}
%Computing the relative ratios between each pair of images can be an exhaustive task, with complexity of \(O(N^2LM)\). In the intermediate regime, a large number of samples is required for an accurate estimate of the signal. A modification of the synchronization algorithm on pairwise ratios must be made in order to process the data in reasonable time. 

\subsubsection{Synchronize and Match}
\revv{Synchronization induces correlation between the shifted signal and the noise because each rotation estimation depends on the measurement's specific noise realization. To weaken the correlation, the key idea of the Synchronize and Match algorithm} is partitioning the data into two subsets of size \(P\) and \(N-P\), with \(P\ll N\). Let us denote these sets by  \(S_P\) and \(S_{N-P}\). Synchronization is applied only to \(S_P\) (the small subset), resulting in \(P\) rotation estimations. 
Next, we assign to each image in \(S_{N-P}\) the rotation assigned to its closest neighbor in \(S_P\) (whose rotation \revv{was} estimated in the previous step).
Finally, in order to break the dependency between the estimated rotations and the noise within \(S_P\), each image in \(S_P\) is assigned with the rotation corresponding to its closest image in \(S_P\), excluding itself. \rev{The algorithm is outlined in Algorithm \ref{SynchAndMatch}.}
\begin{algorithm}
	\caption{Synchronize and Match}
	\label{SynchAndMatch}
	\begin{algorithmic}[1]
		\renewcommand{\algorithmicrequire}{\textbf{Input:}}
		\renewcommand{\algorithmicensure}{\textbf{Output:}}
		\REQUIRE
%		\begin{itemize}
%			\item A set of observations \(y_1,...,y_N\) drawn from (\ref{MRA_model})
%			\item Partition size \(P \ll N\)
%		\end{itemize}
		\rev{A set of observed sPCA coefficients \(v_1,...,v_N\) corresponding to \(Y_1,...,Y_N\)} and partition size \(P \ll N\)
		\ENSURE  \(\hat{\theta}_i \quad \forall i\in[1,N]\)
		
		\STATE Partition the data into two subsets, \(S_P=\{\rev{v}_i\}_{i=1}^P\) and \(S_{N-P}=\{\rev{v}_i\}_{i=P+1}^N\).
		
%		\STATE Compute the relative rotations \(\hat{\theta}_{ij} \quad \forall i,j\in[1,P]\) of the first set of samples \(\{y_i\}_{i=1}^P\).
%		
%		\STATE Estimate the rotations from the relative rotations using PPM, resulting in \(\hat{\phi}_i, \quad \forall i\in[1,P]\)
		
		\STATE Apply synchronization \revv{to} \(S_P\) using the projected power method (see Section \ref{SynchronizationSection}), resulting in a set of rotation estimates \(\hat{\phi}_1,,...,\hat{\phi}_P,\) corresponding to \(\rev{v}_1,,...,\rev{v}_P\).
		
		\FOR {$i = P+1 $ to $N$}
		%			\STATE \(D(n) =  \|y_i - y_n\|_2^2\quad \forall n\in[1,M]\) 
		\STATE \(\hat{\theta}_i = \hat{\phi}_{\arg \min_{n\in[1,P]} {\|\rev{v}_i - \rev{v}_n\|_2^2}}\)
		\ENDFOR
		
		\FOR {$i = 1 $ to $P$}
%		\STATE \(D(n) = \|y_i - y_n\|_2^2 \quad \forall n\in[1,P]\) 
		
%		\STATE \(A(n) = \text{argsort}\{D(),\text{Ascending}\}\)
		
%		\STATE  Compute the indices that would sort the array \(D\) in ascending order, store in array \(A(n), \quad n\in[1,P]\).
%		\STATE \(\hat{\theta}_i = \hat{\phi}_{A(2)}\)
%		\STATE \(j = {\arg \min_{n \in[1,P],n \neq i }
%			\|y_i - y_n\|_2^2}
%		\)
%		\STATE \(\hat{\theta}_i = \hat{\phi}_{\arg \min_{n\in[1,P], n\neq i}
%			\|y_i - y_n\|_2^2}
%			\)
	\STATE \noamrev{\(\hat{\theta}_i = \hat{\phi}_{\arg \min_{n \in[1,P],n \neq i }
		\|v_i - v_n\|_2^2}\)}
		\ENDFOR

%		\RETURN \(\hat{\theta}_i, \quad \forall i\in[1,N]\)
	\end{algorithmic}
\end{algorithm}

The synchronization of the small set has a complexity of \(O(P^2LM)\), while the complexity of assigning rotations to the second set is \(O((N-P)PM)\). 
%The overall complexity is thus \(O(PM(PL+N))\), and for
\revv{Thus, f}or  \tamirrev{$N\gg PL$} the complexity is dominated by \(O(PMN)\).
\nrev{The parameter P controls a trade-off between runtime and accuracy.} \(P\) should be chosen to be large enough such that the distribution of rotations will be concentrated, but should be kept low such that the \nrev{runtime of} Synchronize and Match will be negligible compared to the EM.
%, \(P^2(L-1)M+PMN\ll TMNL\) or \(P^2(L-1)+PN\ll TNL\).

The synchronization step allows us to initialize the EM from the average of the aligned coefficients \(a_0=\frac{1}{N}\sum_{i=1}^N{{v}_i}\), rather than a random guess. This initialization is required for the reduced search space to function properly. Even though EM in general is sensitive to initialization, the initialization by itself does not contribute to the speed up of the EM process. This \nrev{is demonstrated} in \nrev{Figures \ref{results_comparison_N_1000}, \ref{results_comparison_N_5000}, \ref{results_comparison_N_10000}, \ref{results_comparison_N_1000_withsignalprior} and \ref{results_comparison_N_5000_withsignalprior}}, in which standard EM with an initial guess equals to the averaged aligned coefficients \nrev{(denoted as EM+S\&M init)} is compared against standard EM with \nrev{a} random guess. 
%The results prove that the initial guess is not the source of benefit, but enables the use of a reduced search space around the initial guess.

\subsection{Establishing prior on the distribution \rev{of rotations}}
\label{PriorOnDistributionSection}
We are now turning our attention to learning the expected distribution of rotation\rev{s} after synchronization. This will allow us to incorporate this essential information into the EM iterations as a prior, and to narrow down the search space. 
%First, we discuss why this problem is challenging. 
It is important to note that the expected  distribution depends on the specific synchronization method that is being used (in our case, Synchronize and Match), the SNR, and the unknown image. 
%That is why a closed form expression is beyond reach; see Appendix \ref{ClosedFormExpression1d} that demonstrates how even for a toy example of 1-D MRA using template matching with a known reference, an approximation was made for a closed form expression of the expected distribution.
%\\
%\\
\revv{Since it seems {to be} out of reach to compute this distribution analytically}, we introduce a learning methodology to learn the expected distribution of rotations. 
In Appendix~\ref{ClosedFormExpression1d}, we provide \rev{an} approximation for \rev{the expected distribution of 1-D MRA 
using template matching as a synchronization method with a known reference signal.}
\rev{Once the expected distribution is learned, we show how to incorporate it} into the EM iterations as a prior in the Bayesian sense.

\subsubsection{Learning methodology}
\label{DistributionLearningMethodologySection}
Since \rev{synchronization estimate\revv{s}  rotations up} to some global rotation \(\theta_c\), we are interested in the expected distribution of the centered rotation error \(
\Delta\bar{\theta}=\hat{\theta}-\theta-\theta_c\), which can be formulated as:
\begin{equation}
\label{learned_distribution_formula}
p_{\Delta\bar{\theta}}[l]
%=\frac{1}{N}\sum_{j=1}^Np(\Delta\bar{\theta}_j=\Theta_l)
=\mathbb{E}_{\boldsymbol{\theta},\boldsymbol{\varepsilon},x}
\left[\frac{1}{N}\sum_{j=1}^N
\mathcal{I}_{\left\lfloor\frac{\Delta\bar{\theta}_j L}{2\pi}\right\rfloor=l}\right],
\end{equation}
where \(\theta \sim \mathcal{U}\left[0,2\pi\right)\), \(\varepsilon \sim \mathcal{N}(0,\sigma^2I)\), \(x \sim p(x)\)\revvv{,} and~\(\mathcal{I}\) is the indicator function.
%In this section, we provide a method in which the distribution of rotations after synchronization can be learned for a known distribution of signal \(p(x)\) or a fixed signal \(x\), noise variance \(\sigma^2\) and specific synchronization method. 
%In this part we assume that the rotations after synchronization lie on a discrete space \(\Theta\) of length \(L\).
%Since the estimation of the rotations is up to some global rotation \(\theta_c\) per synchronization result, we are interested in the centered rotation error:
%\begin{equation}
%\Delta\bar{\theta}=\hat{\theta}-\theta-\theta_c
%\end{equation}
%The distribution of the centered rotation error is:
%where \(\theta_c\) can be estimated per trial by 
%\begin{equation}
%\label{rotation_center_est}
%\theta_c = \arg\min_{\theta}
%\|R_{\theta}\{\frac{1}{N}\sum_{j=1}^N R_{-\hat{\theta}_j}y_j\}-x\|^2
%\end{equation}
%\begin{equation}
%p_{\Delta\bar{\theta}}[m]=\frac{1}{N}\sum_{j=1}^Np(\hat{\theta}_i-\theta_i=\Theta_m+\theta_c^{(t)})
%\end{equation}
%\begin{equation}
%\label{learned_distribution_formula}
%p_{\Delta\bar{\theta}}[m]=
%
%\end{equation}
The suggested algorithm approximates~\rev{(\ref{learned_distribution_formula})} using Monte-Carlo simulations by computing the sample mean of the measured centered distribution of rotations after synchronization of different trials, for a given noise variance~\(\sigma^2\) and number of observations \(N\).
%\begin{equation}
%p_{\Delta\bar{\theta}}[m]=\frac{1}{N}\sum_{j=1}^N
%\iiint{
%\mathcal{I}_{\Delta\bar{\theta}_i=\Theta_m}|\boldsymbol{\theta},\boldsymbol{\varepsilon}d\boldsymbol{\theta}d\boldsymbol{\varepsilon}}dx
%\end{equation}
The process is detailed in Algorithm \ref{Algorithm 1}.

\begin{algorithm}
	\caption{Algorithm for learning rotation distribution}
	\label{Algorithm 1}
	\begin{algorithmic}[1]
		\renewcommand{\algorithmicrequire}{\textbf{Input:}}
		\renewcommand{\algorithmicensure}{\textbf{Output:}}
		\REQUIRE 
		\rev{
		\begin{itemize}
			\item[]
			
			\item \(p(x)\) - signal prior
			
			\item \(\sigma^2\) - noise variance
			
			\item \(N\) - number of observations

			\item \(f\) - synchronization method
				
			\item \(L\) - number of possible rotations
		
			\item \(R\) - number of repetitions
		\end{itemize}}
%		Signal prior \(p(x)\), noise variance \(\sigma^2\), number of observations \(N\), synchronization method  \(f\), number of possible rotations \(L\), number of repetitions \(R\)
		\ENSURE  Estimated expected centered distribution of rotations after synchronization \(\bar{\rho}[l], l\in[0,L-1]\)
%		\\ \textit{Initialization} :
%		\\ \textit{Main loop}
		\FOR {$r = 1 $ to $R$}
		\STATE Draw \(x\) from \(p(x)\)
		\STATE Generate \(Y_1,...,Y_N\) according to (\ref{MRA_model}), \rev{with \(\theta_1,...,\theta_N\) distributed uniformly over \(\left[0,2\pi\right)\)}
		\STATE Estimate $\hat{\theta}_1,...,\hat{\theta}_N$ using synchronization \(\hat{\boldsymbol{\theta}}=f(y_1,...,y_N)\)
		\STATE Compute the global rotation between the estimated rotations and the signal \\ $\theta_c = \arg\min_{\theta}
		\|R_{\theta}\{\frac{1}{N}\sum_{j=1}^N R_{-\hat{\theta}_j}Y_j\}-x\|^2$
		\STATE Compute the rotation error $\Delta \bar{\theta}_i = \hat{\theta_i}-\theta_i-\theta_c \quad i=0,\ldots,N-1$
		\STATE Compute $\hat{\rho}_{r}[l] =\frac{1}{N} \sum_{i=0}^{N-1}{\mathcal{I}_{\left\lfloor\frac{\Delta\bar{\theta}_j L}{2\pi}\right\rfloor=l}},\quad  l=0,\ldots, L-1$
		%	\STATE Align  $\hat{\rho}[m]$ such that it centered around \(M/2\)
%		\STATE Update $\rho[l] \leftarrow \rho[l] + \frac{1}{R}\hat{\rho}[l]$
		%	\IF {($i \ne 0$)}
		%	\STATE statement..
		%	\ENDIF
		\ENDFOR
		\STATE $\bar{\rho}[l] = \frac{1}{R}\sum_{r=1}^R{\hat{\rho}_{r}[l]} ,\quad  l= 0,\ldots,L-1$
%		\RETURN $\rho[l]$
	\end{algorithmic}
\end{algorithm}

Figure \ref{learned_distribution_2d} shows an example of the learned distribution of rotations {after synchronization}, using the Synchronize and Match algorithm, with \(N=500\), \(L=360\), \nrev{with \(R=50\) repetitions}, and~\(x\) \nrev{is} drawn from a set of \rev{10} projection images of the E. coli 70S ribosome volume taken from \cite{Shaikh2008}.
%(see Section \ref{SynchAndMatchSection}),
Evidently. the learned distribution becomes more concentrated as the SNR increases. This allows us to narrow down the search space of the following EM iterations, \rev{thus reducing its computational load.}

%The step of centering the estimated distribution for each synchronization is necessary since the synchronization estimates the rotations up to a global rotation. Thus, the estimated distribution might be centered differently at each iteration. The alignment can be done by different methods, such as cyclic shifting the distribution by \(\arg\max_m{\hat{\rho}[m]} - \frac{M}{2}\).
%
%\(x\sim\N(0,1)\),\(\sigma2\),\(L=41\),\(N=1000\)
%\begin{figure}[H]
%	\centering
%	\begin{tabular}{@{}c@{}}
%		\includegraphics[width=.7\linewidth,height=75pt]{est_shift_pmf_sigma_5} \\[\abovecaptionskip]
%		\small (a) Template Matching
%	\end{tabular}
%	\begin{tabular}{@{}c@{}}
%		\includegraphics[width=.7\linewidth,height=75pt]{EstPMF_N_1000_sigma_2_L_21.png} \\[\abovecaptionskip]
%		\small (b) Angular Synchronization
%	\end{tabular}	
%	\caption{Estimated distribution of rotations for two different synchronization techniques, template matching and angular synchronization, with \(N=1000\),\(\sigma=2\),\(L=41\) and \(x\sim\ \mathcal{N}(0,1)\).}
%	\label{estimated_dist_example}
%\end{figure}
%\begin{figure}
%	% generated with rotation_err_pmf_verification.m
%	\includegraphics[width=\linewidth]{PaperFigures/LearnedDistribution/mra2d_learned_rotation_dist}
%	\caption{Learned distribution of rotations using Algorithm \ref{Algorithm 1} in different SNRs.
%	}
%	\label{learned_distribution_2d}
%\end{figure}
\noam{
\begin{figure}
	\includegraphics[width=\linewidth]{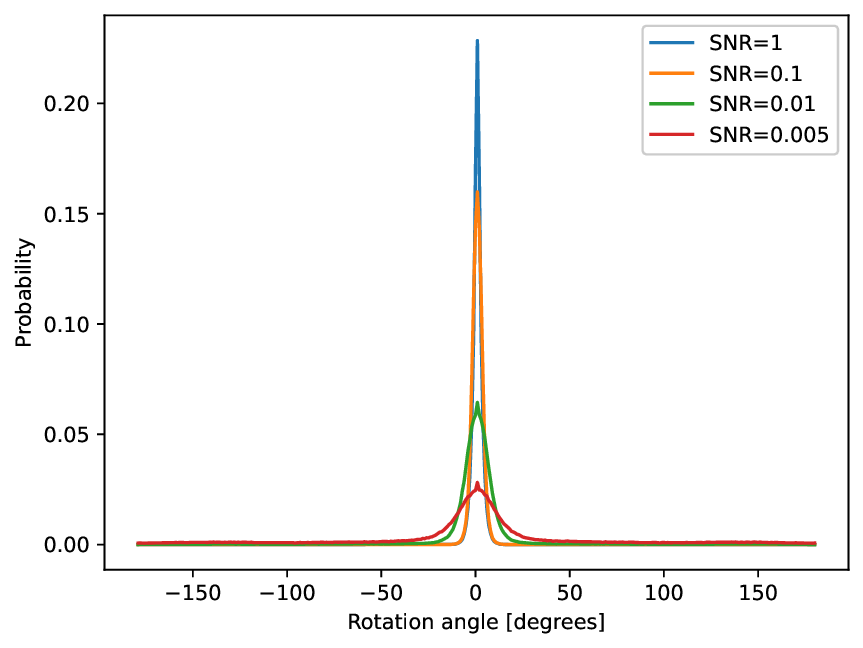}
	\caption{\noam{Learned distribution of rotations using Algorithm \ref{Algorithm 1} in different SNRs.}
	}
	\label{learned_distribution_2d}
\end{figure}}
\subsubsection{Prior on the distribution of rotations}
\label{GibbsPrior}
The prior reflects how we model the distribution of rotations around the expected distribution, \tamirrev{which} was learned using \rev{Algorithm \ref{Algorithm 1}}.
%We suggest a prior that states that a measured distribution is more likely if it is more similar to the expected distribution in the sense of the Kullback–Leibler divergence.
Specifically, once the expected distribution of rotations after synchronization \(\bar{\rho}\) is learned, we assume that the probability of a measured distribution of rotations \(\rho\) is:
\begin{equation}
p(\rho) \propto e^{-\gamma \rev{D_{KL}(\bar{\rho},\rho)}},
\end{equation}
%In this section, we provide different ways to formulate the prior \(p(\rho)\), given the learned distribution of rotations \(\rho_\theta[l]\).
%\subsubsection{Gaussian Prior}
%Assuming the shifts are drawn according to some known distribution \(\bar{\rho}\). The sampled distribution can be thought of as a small perturbation from the known distribution:
%\begin{equation}
%\rho[l] = \bar{\rho}[l] + \epsilon[l]
%\end{equation}  
%where 
%\(\sum_{l=0}^{L-1}{\rho[l]}=1\)
%and
%\(\rho[l] \geq 0 \quad \forall l \in [0,L-1]\).
%Assuming that the perturbation of each element is i.i.d and distributed \(N(0,\sigma_p^2)\), the sampled distribution can be approximated as a Gaussian random vector with distribution:
%\begin{equation}
%p(\rho)=\prod_{l=0}^{L-1}{p(\rho[l])}=
%\frac{1}{(2\pi\sigma_p^2)^{L/2}}e^{-\frac{1}{2\sigma_p^2}\|\rho-\bar{\rho}\|^2}
%\end{equation}
%\subsubsection{Gibbs Prior}
where \(\gamma\) is a regularizer term and
\begin{equation}
D_{KL}(\bar{\rho},\rho) = \sum_{l=0}^{L-1}{\bar{\rho}[l]\log{\frac{\bar{\rho}[l]}{\rho[l]}}}.
\end{equation}
In Appendix \ref{Mstep} we \rev{derive} the update rule of the estimated distribution in {the EM iterations} \revv{(\ref{maximization_step})}
%\revv{step \ref{dist_update} of Algorithm \ref{Synch-EM algorithm}} 
using this prior, \revv{that is used in Algorithm \ref{Synch-EM algorithm}}.
%We use \(E(\rho,\bar{\rho})\) as a measure of similarity between the measured distribution \(\rho\) and the expected distribution \(\bar{\rho}\).
%The regularizer \(\gamma\) controls how similar the measured distribution should be to the expected one for it to be likely.

\subsection{Reduced search space}
\label{ReducedSearchSpaceSection}
%As can be seen from the learned prior on the distribution of rotations after synchronization, the range of possible rotation angles is reduced from uniform range to a smaller range that increases as the SNR decreases. 
As demonstrated in Figure~\ref{learned_distribution_2d},
even in rather low SNR values, the range of likely rotations \rev{remains} small. This allows us to reduce the search space from all possible rotations (because of the uniform distribution of rotations in (\ref{MRA_model})) to a much narrower space of likely rotations after synchronization. 
%Alternatively, the same number of directions can be computed for a smaller range\rev{, increasing the angular} resolution.

Suppose that we consider L equally spaced possible rotations. Namely, the EM iterations search over the angles \(\left[0,\noam{\frac{2\pi}{L}},...,\frac{2\pi(L-1)}{L}\right]\).
If the expected distribution is indeed narrow, it allows us to consider a smaller set of angles 
\noam{\(\left[-\frac{\pi BW}{L},-\frac{\pi BW+2\pi}{L},...,\frac{\pi BW}{L}\right]\)} \rev{for} \(BW\leq L\). \(BW\) should be chosen according to the SNR and the learned distribution. For example, in \rev{the experiment of Figure \ref{learned_distribution_2d} with SNR=1/100\nrev{,} one can set \(BW=36\) with only negligible \revv{loss} of information.}
%The algorithm is similar to the one presented in Section \ref{LowSNR}, with a slight modifications that the E step requires computation of a much smaller set of weights \(\{w_t^{l,j}\}_{l=0, j=1}^{l=BW-1,j=N}\), where \(w_t^{l,j}\) is computed as before according to (\ref{weights_2d_em}).
%The maximization would then be:
%\begin{equation}
%a_{t+1} =
%\frac{1}{N}\sum_{j=1}^{N}{\sum_{l=0}^{BW-1}
%	{
%		w_{t}^{l,j}e^{\frac{2\pi \I \bar{k}l}{L}}\circ v_j
%	}
%}
%\end{equation}
%and maximization over the distribution:
%\begin{equation}
%\rho_{t+1}[l]=\frac{W_t[l]+\gamma\overline{\rho}[l]}{\sum_{l=0}^{BW-1}(W_t[l]+\gamma\overline{\rho}[l])} \quad \forall l\in[0,BW-1]
%\end{equation}
%\\
%\\
Similarly, one can retain the same computational load while increasing resolution by sampling the range of likely rotations more densely \secrev{\(\left[-\frac{\pi BW}{L},-\frac{\pi(L-1)BW}{L^2},...,\frac{\pi(L-1)BW}{L^2},\frac{\pi BW}{L}\right]\)}, with the appropriate modification of the EM.
%
%Alternatively, the continuous space \(SO(2)\) can be still sampled in L directions in a smaller range of angles \(\left[0,\frac{2\pi BW}{L^2},...\frac{2\pi(L-1)BW}{L^2}\right]\) which results in improved resolution. The E step would then require computation of  \(\{w_t^{l,j}\}_{l=0, j=1}^{l=L-1,j=N}\), where
%\begin{equation}
%w_t^{l,j} =
%C_t^j
%\exp{
%	\left(
%	-\frac{1}{2\sigma^2}\|e^{\frac{2\pi \I  BW\bar{k}l}{L^2}}\circ a_t-v_j \|_2^2
%	\right)
%} \rho_{t}[l]
%\end{equation} 
%And the maximization differs in
%\begin{equation}
%a_{t+1} =
%\frac{1}{N}\sum_{j=1}^{N}{\sum_{l=0}^{L-1}
%	{
%		w_{t}^{l,j}e^{\frac{2\pi \I  BW\bar{k}l}{L^2}}\circ v_j
%	}
%}
%\end{equation}
%while maximization over the distribution:
%\begin{equation}
%\rho_{t+1}[l]=\frac{W_t[l]+\gamma\overline{\rho}[l]}{\sum_{l=0}^{L-1}(W_t[l]+\gamma\overline{\rho}[l])} \quad \forall l\in[0,L-1]
%\end{equation}

%\section{Method Of Moments}
\section{Numerical {Experiments}}
In this section, we compare Algorithm \ref{Synch-EM algorithm} to standard EM \revv{applied to the statistical model (\ref{MRA_model})} (see, for example, \cite{BispectrumInversion}) in different scenarios. We use projection images of the E. coli 70S ribosome volume taken from \cite{Shaikh2008}. Each image is of size \(129\times 129\) pixels. The code to reproduce all experiments (including those presented in previous sections) is publicly available at \url{https://github.com/noamjanco/synch-em}.

\subsection{Performance without signal prior}
\begin{figure}
	\includegraphics[width=\linewidth]{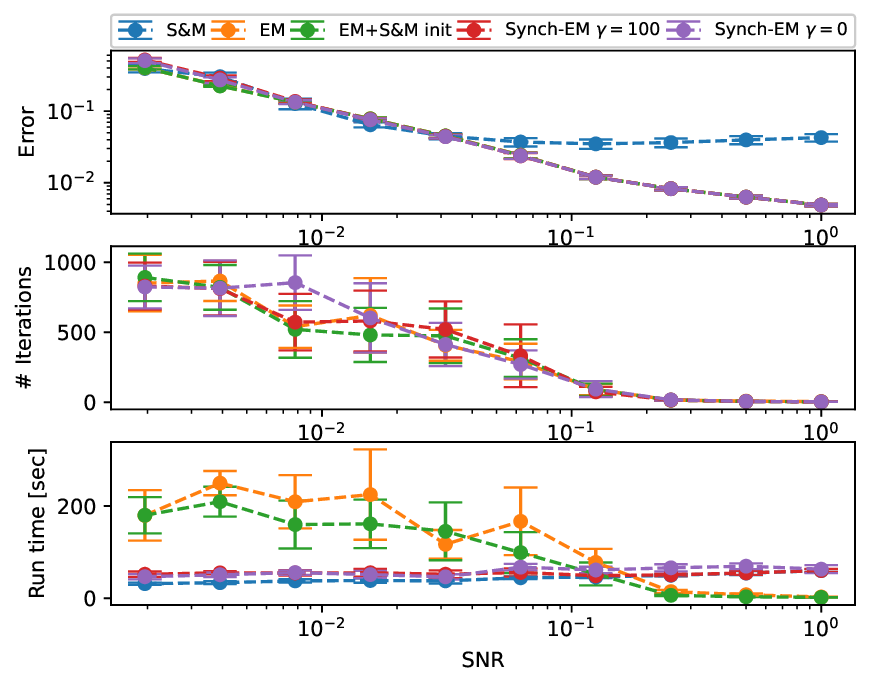}
%	\caption{Performance comparison of standard EM, Synchronize and Match, and Synch-EM, with \(N=1000\).	}
	\caption{Performance of Synchronize and Match (S\&M), standard EM with random initialization (EM), standard EM initialized from the average of the aligned coefficients after S\&M (EM+S\&M init), and Synch-EM with different distribution prior weighting (Synch-EM \(\gamma=100\) and \(\gamma=0\)), as a function of the SNR, with \(N=1000\), \secrev{averaged over 10 trials.}}
	\label{results_comparison_N_1000}
\end{figure}

\begin{figure}
	\includegraphics[width=\linewidth]{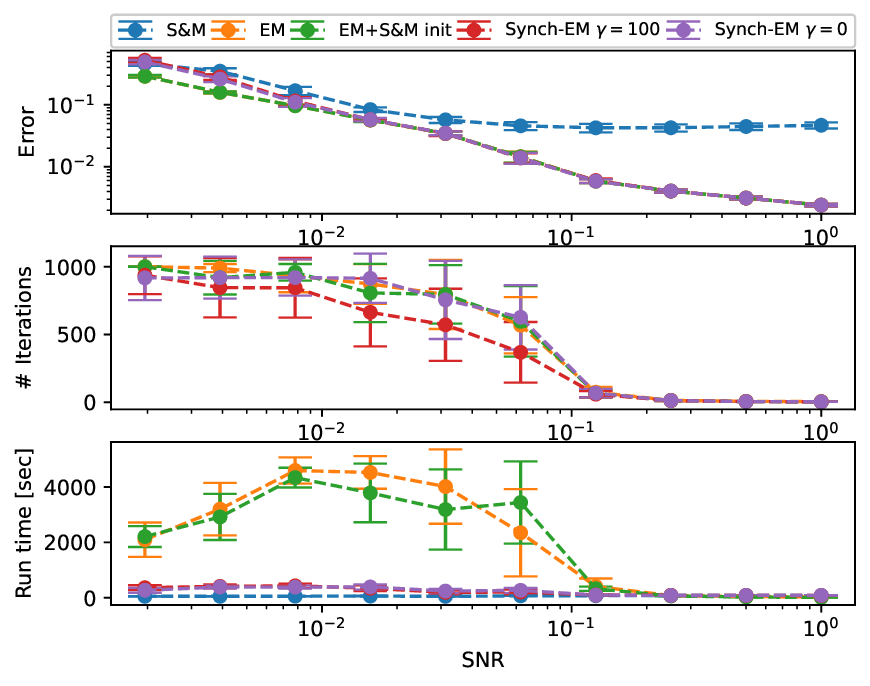}
%	\caption{\noam{Performance comparison of standard EM, Synchronize and Match, and Synch-EM, with \(N=5000\).}
%	}
	\caption{\nrevv{Performance of Synchronize and Match (S\&M), standard EM with random initialization (EM), standard EM initialized from the average of the aligned coefficients after S\&M (EM+S\&M init), and Synch-EM with different distribution prior weighting (Synch-EM \(\gamma=100\) and \(\gamma=0\)), as a function of the SNR, with \(N=5000\)}, \secrev{averaged over 10 trials.}}
	\label{results_comparison_N_5000}
\end{figure}

\begin{figure}
	\includegraphics[width=\linewidth]{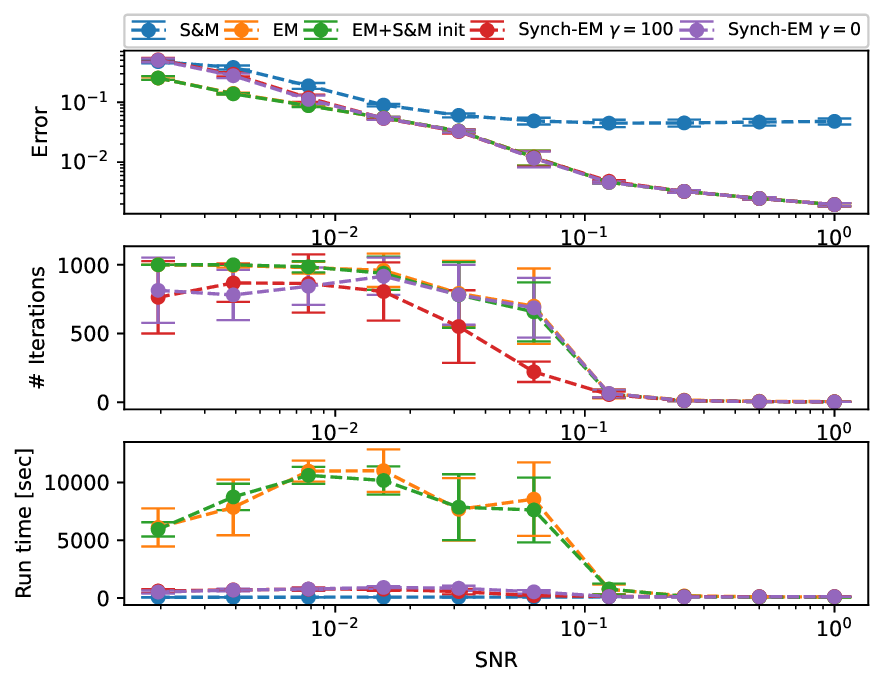}
%	\caption{\noam{Performance comparison of standard EM, Synchronize and Match, and Synch-EM, with \(N=10000\).}
%	}
	\caption{\nrevv{Performance of Synchronize and Match (S\&M), standard EM with random initialization (EM), standard EM initialized from the average of the aligned coefficients after S\&M (EM+S\&M init), and Synch-EM with different distribution prior weighting (Synch-EM \(\gamma=100\) and \(\gamma=0\)), as a function of the SNR, with \(N=10000\)}, \secrev{averaged over 10 trials.}}
	\label{results_comparison_N_10000}
\end{figure}

We first examine the \rev{Synch-EM} \revv{algorithm} without taking into account the signal prior \(p(a)\) within the EM iterations.
%, but only using the learned distribution of rotations after Synchronize and Match (Algorithm \ref{Algorithm 1}). 
Figures \ref{results_comparison_N_1000}, \ref{results_comparison_N_5000} and \ref{results_comparison_N_10000} compare the relative error, number \rev{of} iterations and runtime of Synchronize and Match, standard EM with random initialization, standard EM initialized from the average of the aligned coefficients after \nrevv{Synchronize and Match}, and Synch-EM with different distribution prior weighting, as a function of the SNR, for \(N=1000\), \(N=5000\) and \(N=10000\), respectively. For each SNR, the results were averaged over 10 trials. \rev{W}e used \(L=360\), \(P=100\), \(\gamma=100\), a fixed \(BW=36\), \(T=1000\), and \(tol=10^{-5}\). \revv{The runtime of Synch-EM includes Synchronize and Match.} 
%\noam{The figures also show the standard deviation using the error bar}.
\nrev{Table~\ref{table:7} also presents a runtime comparison for \(N=10000\) for the SNRs of interest.}
\rev{There is a clear and significant improvement in runtime of Synch-EM compared to standard EM in all scenarios, \revv{which gets more \tamirrev{significant} as \(N\) increases.}}
The improvement \rev{in} runtime \rev{results from two compounding \tamirrev{affects}:} the computational burden of each EM iteration is reduced due to the {narrower} search space after synchronization, and the {reduced} number of EM iterations (\revv{for \(N=5000\) and \(N=10000\)}) in the \rev{low SNRs (\revv{lower than 1/10})}.
\revv{We see that as \(N\) increases, the \revv{reduction} in the number of iterations is more significant.} 
\rev{For example, for \(N=10000\) (Figure \ref{results_comparison_N_10000}) and \(\text{SNR}=1/16\), the average number of iterations using standard EM was 699, compared to 221 using  Synch-EM. The average \revv{running} time using standard EM was 7064 seconds, compared to 166 seconds using  Synch-EM.}
We {also} see that the choice of \(BW=36\) \revv{has a negligible effect on the accuracy, unless the SNR is below 1/100.}
%\begin{figure}[H]
%	\includegraphics[width=\linewidth]{PaperFigures/Results_WithoutSignalPrior/Synchronize_and_Match_EM_Synch-EM_N_1000_sPca_1}
%	\caption{Performance comparison of standard EM, Synchronize and Match, and Synch-EM, with \(N=1000\).
%	}
%	\label{results_comparison_N_1000}
%\end{figure}

%\begin{figure}
%	\includegraphics[width=\linewidth]{PaperFigures/Results_WithoutSignalPrior/Synchronize_and_Match_EM_Synch-EM_N_1000_sPca_1}
%	\caption{Performance comparison of standard EM, Synchronize and Match, and Synch-EM, with \(N=1000\).
%	}
%	\label{results_comparison_N_1000}
%\end{figure}
%\begin{figure}
%	\includegraphics[width=\linewidth]{PaperFigures/Results_WithoutSignalPrior/Synchronize_and_Match_EM_Synch-EM_N_5000_sPca_1}
%	\caption{Performance comparison of standard EM, Synchronize and Match, and Synch-EM, with \(N=5000\).
%	}
%	\label{results_comparison_N_5000}
%\end{figure}
%\begin{figure}
%	\includegraphics[width=\linewidth]{PaperFigures/Results_WithoutSignalPrior/Synchronize_and_Match_EM_Synch-EM_N_10000_sPca_1}
%	\caption{Performance comparison of standard EM, Synchronize and Match, and Synch-EM, with \(N=10000\).
%	}
%	\label{results_comparison_N_10000}
%\end{figure}


\begin{table}[h!]
	\centering
		\begin{tabular}{ |c| c c c c| } 
			\hline
			SNR & \(1/8\) & \(1/16\) & \(1/32\) & \(1/64\)\\ 
			\hline
			Synchronize \nrevv{and} Match & 93 & 85  & 74 & 73\\ 
			
			EM & 746 & 8564 & 7681 & 11031 \\ 
			%EM + init  & cell5 & cell6 & cell5 & cell6 \\ 

			EM + S\&M init & 776 & 7624  & 7865 & 10179\\ 
			
			Synch-EM \(\gamma=100\)& 133 & \textbf{205}  & \textbf{552} & \textbf{759}\\ 
			
			Synch-EM \(\gamma=0\) & \textbf{127} & 534  & 844 & 913 \\
			\hline
		\end{tabular}
		
%		\caption{Run time comparison of standard EM, Synchronize and Match, and Synch-EM, with \(N=10000\).}
	\caption{\nrevv{Runtime in seconds comparison of Synchronize and Match, standard EM with random initialization (EM), standard EM initialized from the average of the aligned coefficients after S\&M (EM+S\&M init), and Synch-EM with different distribution prior weighting (Synch-EM \(\gamma=100\) and \(\gamma=0\)), as a function of the SNR, with \(N=10000\)}.}
		\label{table:7}
\end{table}
\subsection{Performance with signal prior}
\begin{figure}[H]
	\includegraphics[width=\linewidth]{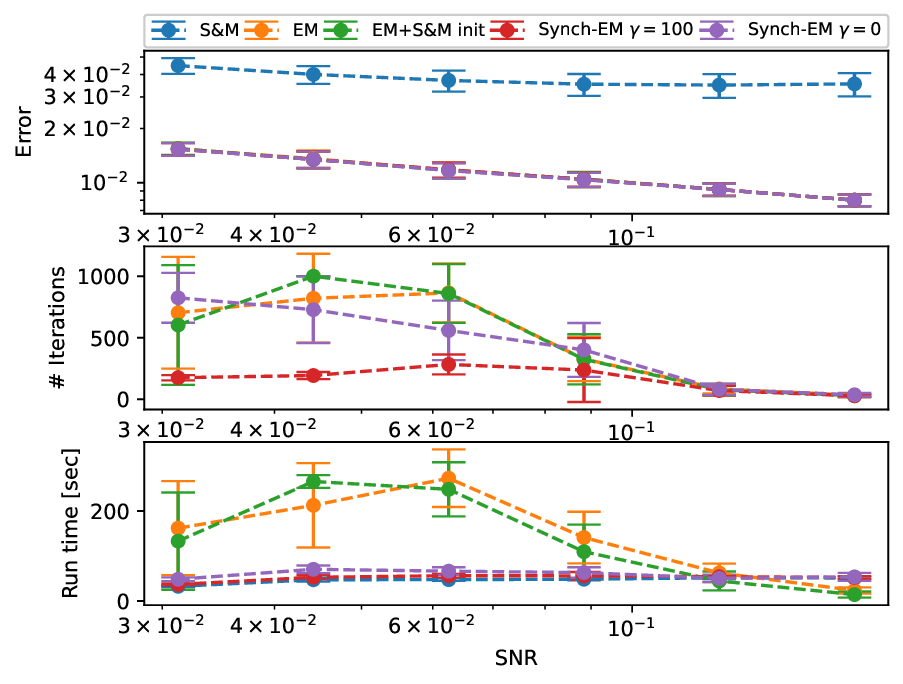}
%	\caption{\noam{Performance comparison of standard EM, Synchronize and Match, and Synch-EM, with \(N=1000\) \revv{and} a signal prior.}
%	}
	\caption{\nrevv{Performance of Synchronize and Match (S\&M), standard EM with random initialization (EM), standard EM initialized from the average of the aligned coefficients after S\&M (EM+S\&M init), and Synch-EM with different distribution prior weighting (Synch-EM \(\gamma=100\) and \(\gamma=0\)), as a function of the SNR, with \(N=1000\) and a signal prior, \secrev{averaged over 10 trials.}}}
	\label{results_comparison_N_1000_withsignalprior}
\end{figure}

\begin{figure}[H]
	\includegraphics[width=\linewidth]{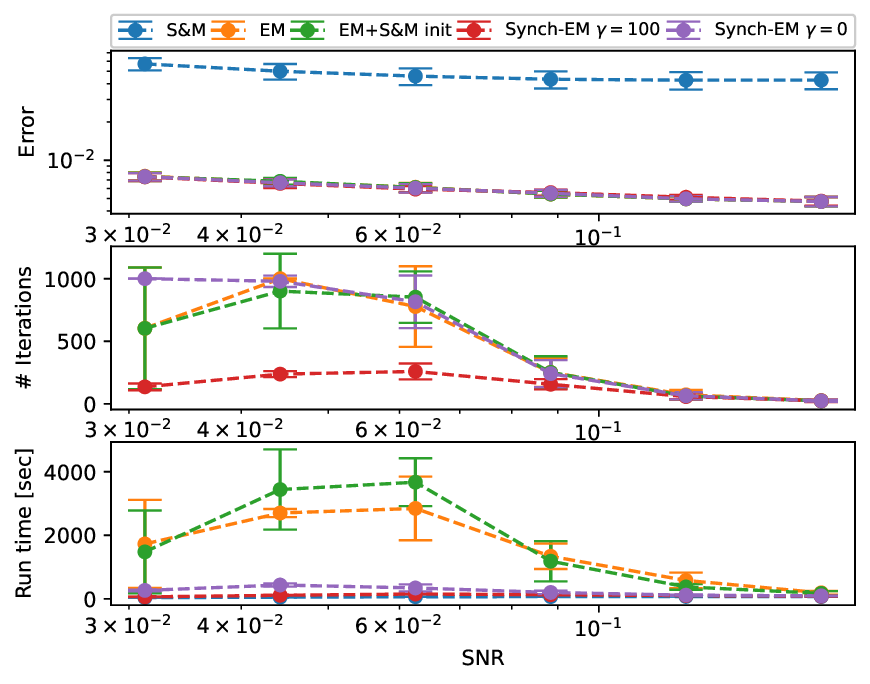}
%	\caption{\noam{Performance comparison of standard EM, Synchronize and Match, and Synch-EM, with \(N=5000\) \revv{and} a signal prior.}
%	}
	\caption{\nrevv{Performance of Synchronize and Match (S\&M), standard EM with random initialization (EM), standard EM initialized from the average of the aligned coefficients after S\&M (EM+S\&M init), and Synch-EM with different distribution prior weighting (Synch-EM \(\gamma=100\) and \(\gamma=0\)), as a function of the SNR, with \(N=5000\) and a signal prior, \secrev{averaged over 10 trials.}}}
	\label{results_comparison_N_5000_withsignalprior}
\end{figure}

We repeated the same experiment with an explicit signal prior \(p(a)\) \revv{incorporated into the EM iterations, in both standard EM and Synch-EM}.
%In this experiment, the algorithm explicitly uses the signal prior \(p(a)\) and the learned distribution of rotations after Synchronize and Match.
The signal prior assumes that \(a\sim\mathcal{CN}(0,\Gamma_a)\), \revv{where \(\Gamma_a\) is a diagonal {covariance} matrix whose entries decay exponentially {with} the angular frequency of the sPCA coefficient: \(\Gamma_a=\text{diag}((4e^{-\bar{k}/8})^2)\)}.
% \(\Gamma_a=\text{diag}((4e^{-\bar{k}/8})^2)\). 
%The covariance assumes exponential decay model of the sPCA coefficients in the angular frequencies.
The decay parameters were \rev{empirically} fitted \rev{\revv{based on} 20 clean projection images of the E. coli 70S ribosome volume taken from \cite{Shaikh2008}}.
Figures \ref{results_comparison_N_1000_withsignalprior} and~\ref{results_comparison_N_5000_withsignalprior} \rev{present} the relative error, number of iterations and runtime of standard EM, Synchronize and Match, and Synch-EM, for \(N=1000\) and \(N=5000\), respectively. 
\nrev{Table \ref{table:13} also presents a runtime comparison for \(N=5000\) for the SNRs of interest.}
\revv{We notice that the signal prior improves the relative error of both standard EM and Synch-EM, which outperform synchronization ({compare, for example,} SNR=0.03 in Figure~\ref{results_comparison_N_1000} and Figure~\ref{results_comparison_N_1000_withsignalprior}). The error of both algorithms remains quite low in low SNRs}.
Again, we see a significant improvement in runtime due to the reduced number of iterations and the \tamirrev{narrow} search space in the SNRs of interest (\revv{below} 1/10). %\marginnote{I think runtime means the instance of time in which a program is running, and not the amount of time}
%With the assumption that \(a\sim\mathcal{CN}(0,\Gamma_a)\), the covariance was where \(\Gamma_a=\text{diag}((4e^{-\bar{k}/8})^2)\) was fitted using regression.
%The prior on the distribution of the signal \(p(a)\) is complex normal with a fitted covariance \(\Gamma_a=\text{diag}((4e^{-\bar{k}/8})^2)\).
%\begin{figure}[h]
%	\includegraphics[width=\linewidth]{PaperFigures/Results_WithSignalPrior/Synchronize_and_Match_EM_Synch-EM_N_1000_sPca_1}
%	\caption{Performance comparison of standard EM, Synchronize and Match, and Synch-EM, with \(N=1000\) \revv{and} a signal prior.
%	}
%	\label{results_comparison_N_1000_withsignalprior}
%\end{figure}

\begin{table}[h!]
	\centering
		\begin{tabular}{ |c| c c c c| } 
			\hline
			SNR & \(0.125\) & \(0.0883\) & \(0.0625\) & \(0.0441\)\\ 
			\hline
			Synchronize \nrevv{and} Match & 74 & 69  & 65 & 51\\ 
			
			EM & 579 & 1339 & 2844 & 2699 \\ 
			%EM + init  & cell5 & cell6 & cell5 & cell6 \\ 

			EM + S\&M init & 375 & 1183  & 3669 & 3437\\ 
			
			Synch-EM \(\gamma=100\)& \textbf{102} & \textbf{128}  & \textbf{151} & \textbf{112}\\ 
			
			Synch-EM \(\gamma=0\) & 115 & 200  & 342 & 435 \\
			\hline
		\end{tabular}
		
%		\caption{Run time comparison of standard EM, Synchronize and Match, and Synch-EM, with \(N=5000\) and signal prior.}
		\caption{\nrevv{Runtime in seconds comparison of Synchronize and Match, standard EM with random initialization (EM), standard EM initialized from the average of the aligned coefficients after S\&M (EM+S\&M init), and Synch-EM with different distribution prior weighting (Synch-EM \(\gamma=100\) and \(\gamma=0\)), as a function of the SNR, with \(N=5000\) and a signal prior}.}
		\label{table:13}
\end{table}

\begin{figure}[H]
	\includegraphics[width=\linewidth]{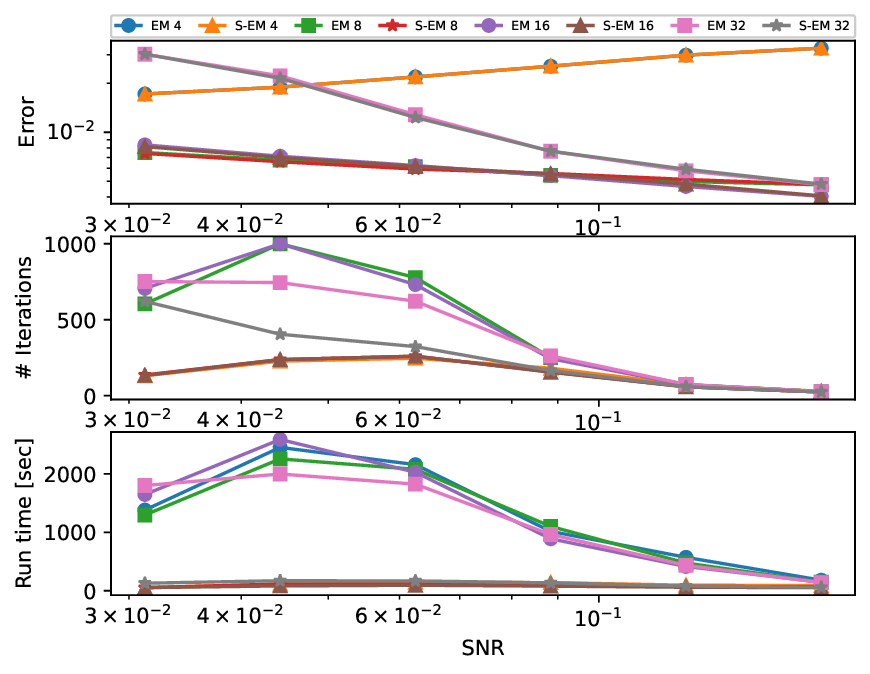}
	\caption{\nrev{Performance comparison between standard EM (EM) and Synch-EM (S-EM) over different choices of signal prior. The label describes the decay parameter \(\beta\) in the assumed covariance of signal's coefficients \(\Gamma_a=\text{diag}((4e^{-\bar{k}/\beta})^2)\), with \(N=5000\) observations. For example, EM~4 refers to standard EM with \(\beta=4\) and S-EM~4 refers to Synch-EM with \(\beta=4\), \secrev{averaged over 10 trials.}
	}}
	\label{signal_prior_choices_5000}
\end{figure}

The choice of the explicit signal prior affects both standard EM and Synch-EM. In the following experiment, we evaluated the relative error, number of iterations and runtime of standard EM and Synch-EM, for different choices of the decay parameter \(\beta\) in the assumed signal coefficient's covariance \(\Gamma_a=\text{diag}((4e^{-\bar{k}/\beta})^2)\).
%, in order to see how a poor choice of signal prior affects the performance. 
Figure \ref{signal_prior_choices_5000} shows the results of the experiment, \nrevv{with} \(N=5000\) observations and 10 trials for each SNR \nrevv{level}. The experiment verifies our choice of \(\beta=8\), \nrevv{and} also shows that the choice of the signal prior has a similar effect on the relative error performance of both the standard EM and the Synch-EM.

\subsection{Prior \revv{weighting}}
\begin{figure}[h!]
	\includegraphics[width=\linewidth]{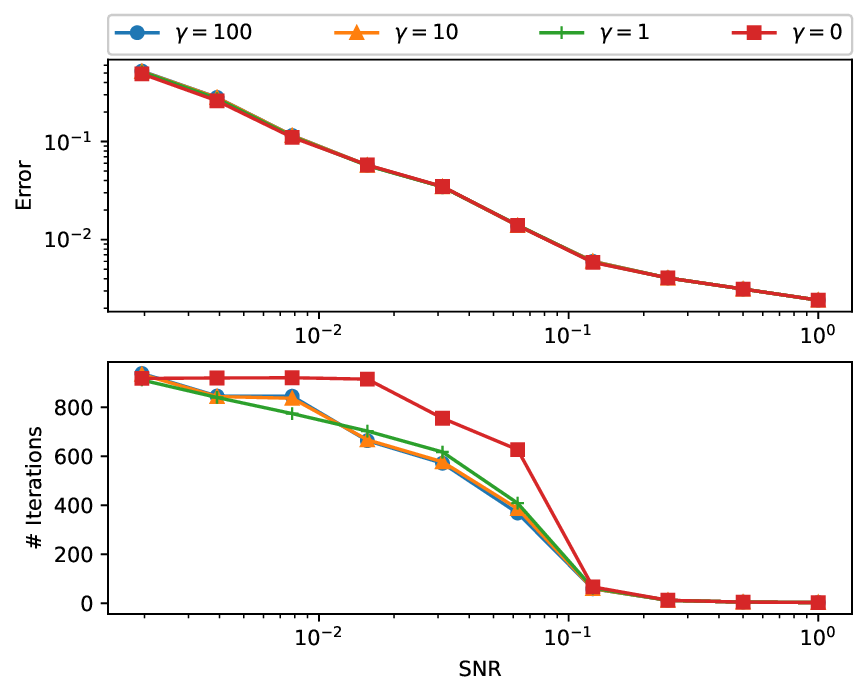}
	\caption{Performance of Synch-EM {with} different values of \tamirrev{$\gamma$: the weight of the prior on the distribution of rotations.} We see that non-zero weight reduces the number of iterations in low SNR.
	}
	\label{results_regularizer}
\end{figure}

In \nrev{this} experiment, we demonstrate how different values of \(\gamma\), the \rev{weight} of the prior on the distribution of rotations after synchronization, affect the EM. % in terms of relative error and number of iterations. 
In this experiment we used \(L=360\), \(P=100\), \(BW=36\), \(T=1000\), \(tol=10^{-5}\), \(N=5000\), and no explicit \revv{signal prior}. 
Figure~\ref{results_regularizer} shows the relative error and number iterations of Synch-EM as a function of SNR, for \(\gamma=0,1,10,100\). 
For each SNR, the results were averaged over 10 trials.
The case of \(\gamma=0\) corresponds to having no prior on the distribution of rotations after synchronization (besides the reduced search space).
As can be seen, the prior reduces the number of EM iterations for values greater than 0, \revv{for low SNR values}, without increasing the relative error. 

\subsection{Performance with different BW}
\begin{figure}[h!]
	\includegraphics[width=\linewidth]{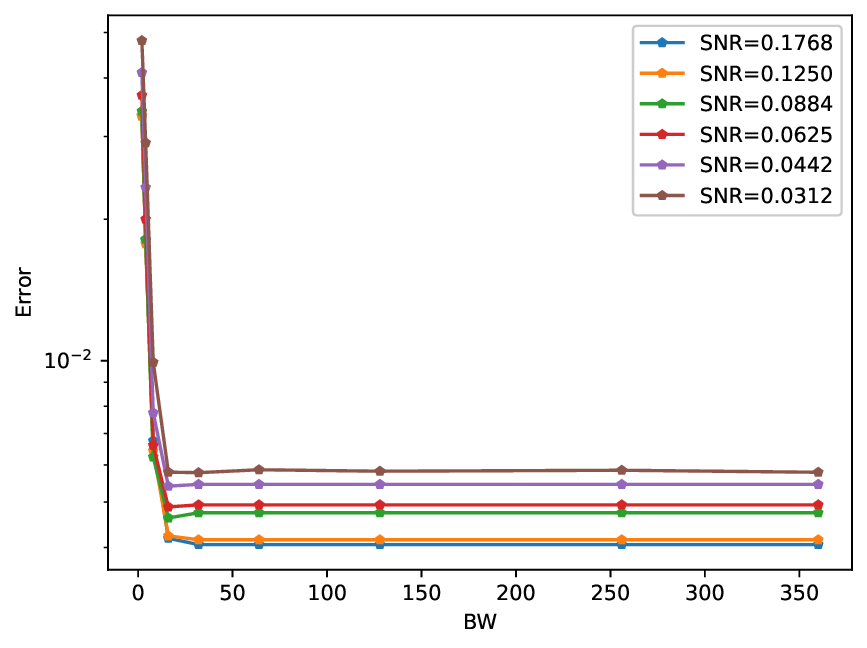}
	\caption{\nrev{Error as a function of the parameter \(BW\) with signal prior for different SNR levels.}}
	\label{err_vs_bw2}
\end{figure}

In \nrev{the last} experiment we tested different choices of BW on the relative error performance.
Figure \ref{err_vs_bw2} shows the relative error as a function of different \(BW\) values with \(N=10000\) samples, with signal prior, for different SNR levels. 
\nrev{The experiment indicates that the error does not decrease for BW greater than 32.} 
Another interesting phenomena is that for specific \nrev{working regimes} (for example \(\text{SNR}=0.0884\), \(BW=16\)) we get a \nrevv{slight} improvement in the relative error. This improvement can be explained by fact that the reduced search space acts as a smoothing prior since it does not allow rotations far from the initial guess.
%, which results in a smoother image that sometimes better matches the true underlying image.

\section{Conclusions and \revv{potential} implications to cryo-EM}
In this paper, we have presented the Synch-EM framework, {where} a modified synchronization algorithm is incorporated into EM. This framework introduce{s} three features that {provide} \revv{significant runtime acceleration}. 
First, 
\tamirrev{the synchronization algorithm approximates the image, and this approximation} %\revv{{results in} an approximation of the image
can be used to initialize the EM. Second, it allows establishing a prior on the distribution of rotations after synchronization.
Third, since the distribution of rotations after synchronization tends to be concentrated (unless the SNR is very low), the search space can be {narrowed down}, thus reducing the computational burden.
In the SNR range between~1/100 to~1/10, we achieve a significant improvement in runtime, \rev{without compromising accuracy}. 

%\begin{figure}[t]
%	\includegraphics[width=\linewidth]{PaperFigures/Results_WithSignalPrior/Synchronize_and_Match_EM_Synch-EM_N_5000_sPca_1}
%	\caption{Performance comparison of standard EM, Synchronize and Match, and Synch-EM, with \(N=5000\) \revv{and} a signal prior.
%	}
%	\label{results_comparison_N_5000_withsignalprior}
%\end{figure}

\revv{The recent interest in the MRA model, and this paper in specific,} \tamirrev{is} mainly motivated by the emerging technology of cryo-EM to reconstruct 3-D molecular structures \cite{bendory2020single}. We now outline how the Sync-EM framework can be extended, in principle, to cryo-EM \revv{datasets}. 
%In cryo-EM, \(\phi:\mathbb{R}^3\to\mathbb{R}\) represents the 3-D molecular structure to be estimated.
Under some simplifying assumptions, {a} cryo-EM experiment results in a series of images, modeled as 
\begin{equation}
\label{cryoEM_model}
I_i = {h} \ast T_{t_i}PR_{\omega_i}\phi +\varepsilon_i, \quad i=1,...,N\rev{,}
\end{equation}
where \(\phi:\mathbb{R}^3\to\mathbb{R}\) 
is a representation of the 3-D structure to be recovered, \(R_{\omega}\) is \revv{a 3-D rotation by} \revv{\(\omega\in\text{SO}(3)\)}, P is a fixed tomographic projection, \(T_t\) is a 2-D shift \revv{by \(t\in \mathbb{R}^2\)}, \(h\) is the point spread function of the microscope, and \(\varepsilon \sim\ \mathcal{N}(0,\sigma^2I)\) is \rev{an additive} noise.
%which represents the 3-D molecular structure to be estimated, by a 3-D rotationy \(R_{\omega}\), integrating along the z-axis (tomographic projection), shifting by a 2-D shift \(T_t\), convolving with the point spread function of the microscope \(h\) and adding noise \(\varepsilon_i \sim\ \mathcal{N}(0,\sigma^2I)\).
%where \(G\) is the group of 3-D rotations and T is a linear operator that takes rotated structure, integrates along the z-axis, convolves with the point spread function and samples it on a Cartesian grid, and \(\varepsilon_i \sim\ \mathcal{N}(0,\sigma^2I)\).
%In cryo-EM, there is no informative prior on the distribution of rotations over the group of 3-D rotations.
\revv{To estimate \(\phi\) from the images \(I_1,\ldots,I_N\), popular software packages \tamirrev{implement} the EM algorithm~\cite{scheres2012relion,punjani2017cryosparc}.}
%The most successful cryo-EM software packages are aiming at estimating \(\phi \) using EM~\cite{scheres2012relion,punjani2017cryosparc}.
Notably, \revv{all packages in the field} assume a uniform prior over the distribution of 3-D rotations.
%The signal \(x \in \mathbb{R}^L\) is assumed to lie in a known space, such as the space of signals with a finite spectral expansion, that random elements of \(G\) are act upon. The task is to estimate the signal \(x\) from the set of observations \(\{y_i\}_{i=1}^N\), while the group elements \(\{g_i\}_{i=1}^N\) are unknown. The recovery is possible up to left multiplication by some group element \(g\in G\), which means we wish to estimate the orbit of \(x\) under group action. 

Similarly to the 2-D MRA model (\ref{MRA_model}), there is an established technique to estimate the 3-D rotations from cryo-EM images~\cite{singer2011three}.
However, since \revv{it is impossible to} estimate rotations accurately in low SNR (the SNR in cryo-EM datasets can be as low as 1/100), it is used only to construct low-resolution ab initio models \cite{greenberg2017common,bandeira2020non}.
A modified version of Algorithm \ref{Synch-EM algorithm} can be \revv{designed} to learn the expected distribution after synchronization, and to narrow down the search space. Importantly, since EM for cryo-EM samples the group of 3-D rotations SO(3) (rather than SO(2) in (\ref{MRA_model})), restricting the search space might lead to dramatic acceleration, far exceeding the acceleration factor presented in this paper. \revv{In addition,} Sync-EM can be combined with other techniques which are currently used to narrow the search space along the EM iterations~\cite{scheres2012relion,punjani2017cryosparc}.

\bibliographystyle{plain}
%\bibliography{research_article.bbl}
%\bibliography{bibfile_article_2}
%\input{research_article.bbl}

\appendices

\section{\revv{Fourier-Bessel \tamirrev{Basis} and Steerable PCA}}
\label{FBCoeffs}
\revv{Before {applying steerable PCA (sPCA)}, we expand all observable images \(Y_1,...,Y_N\) in Fourier-Bessel basis, which is a steerable basis.}
The Fourier-Bessel \revv{basis} on a disk with radius \(c\) is defined as:
\begin{equation}
u^{k,q}(r,\theta)=
\begin{cases}
N_{k,q}J_k(R_{k,q} \frac{r}{c})e^{\I k\theta}, & r \leq c \\
0, & r > c
\end{cases},
\end{equation}
where \(J_k\) is the Bessel function of the first kind, \(R_{k,q}\) is the \(q^{th}\) root of \(J_k\) and \(
N_{k,q}=(c\sqrt{\pi}|J_{k+1}(R_{k,q})|)^{-1}
\) is a normalization factor \cite{2dMRA}, \cite{zhao2016fast}. We take \(c\) to be \((L-1)/2\) according to the support of the images.

%Due to the Nyquist sampling criterion, the Fourier-Bessel expansion requires components for which \(R_{k,q}/2\pi\leq L/4\). 
We assume that the true image can be represented accurately using these components:
\begin{equation}
\label{fb_expansion2}
I(r,\theta)=\sum_{k=-L}^L\sum_{q=1}^{p_k}{a_{k,q}u^{k,q}(r,\theta)},
\end{equation}
where \(p_k\) denotes the number of components, determined by \nrevv{a} sampling criterion (the analog of Nyquist sampling Theorem) that dictates \(R_{k,q}\leq \pi L/2\) \cite{zhao2013fourier}.

The process of computing the Fourier-Bessel coefficients \cite{zhao2016fast} is done first by
sampling the continuous Fourier-Bessel basis per pixel within the circle of radius \(L/2\) from the center of the image:
\begin{equation}
U^{k,q}(x,y) = u^{k,q}(r(x,y),\theta(x,y)).
\end{equation}
The matrix \(U^{k,q}\) per pixel is rearranged as a vector, and each vector per pixel is placed as a row in a matrix \revv{\(\Psi'\in \mathbb{C}^{N\times M'}\)}, where \(N\) is the number of pixels within the circle, and \revv{\(M'\)} is the number of pairs of angular and radial frequencies used in the representation. Given an image in the pixel space \(I\) and using (\ref{fb_expansion}), the Fourier Bessel coefficients are computed \rev{as} the least square solution:
\(\alpha^*=\min_\alpha{\|\Psi' \alpha - I\|_2^2}=(\Psi'^*\Psi')^{-1}\Psi'^* I\).
\rev{Note that} \(\Psi'^\dagger=(\Psi'^*\Psi')^{-1}\Psi'^*\) can be computed once and then applied to the observations (in parallel).
\nrev{We also note that although} \noam{the Fourier-Bessel functions are orthogonal as continuous functions, their discrete sampled versions on Cartesian grid are not necessarily orthogonal. However, they approach orthogonality as \(L\to\infty\) \cite{zhao2013fourier}. Numerical observations suggest that for moderate values of \(L\) this approximation holds.} \nrev{This means that the white noise remains approximately white after the transformation, which allows us to reformulate the problem as in (\ref{MRA_Model_FB}).}

\revv{Then, to reduce the dimensionality of the representation and to denoise the image, {sPCA} is applied \cite{zhao2016fast}. The sPCA results in a new data driven basis \revv{\(\Psi\in \mathbb{C}^{N\times M}\)} to represent the images, while preserving the steerability property. The number of sPCA coefficients being used \(M<M'\) is automatically chosen as a function of the SNR~{\cite{zhao2016fast}}. Finally, all observable images \(Y_1,...,Y_N\) are expanded in the sPCA basis, which results in a set of observable coefficients \(v_1,...,v_N\), where \(v_i\in\mathbb{C}^M\).} %The vectors \(\bar{k}, \bar{q} \in \mathbb{Z}^M\) specify the angular and radial \revvv{indices} at each entry of the coefficient vector.}
%The Fourier-Bessel coefficients are computed per observation by computing the pseudo inverse of \(\Psi\) once and multiplying by the image vector.

\section{\revv{An approximation of} the Distribution of Shifts in a toy problem}
\label{ClosedFormExpression1d}
\rev{We are interested in the error distribution of the estimated shifts in 1-D MRA, where the shift estimation is performed using template matching \revv{with} the true signal as the reference signal. We assume \(x\in\mathbb{R}^L\) drawn from \(\mathcal{N}(0,I)\).}
%Consider the problem of estimating the cyclic shifts of 1-D MRA observations with a known signal \(x\in\mathbb{R}^L\) drawn from \(\mathcal{N}(0,I)\) using template matching as in (\ref{MLE_shift}). 
\revv{To ease notation, we} consider the case where the MRA observations have zero shift.
% which allows us to consider the estimated shift as the shift estimation error and simplifies notation.
We wish to estimate the probability mass function \(p_{\theta}[m]=p(\theta=m)\) of the discrete random variable \(\theta\), defined by:
\begin{equation}
\label{MLE_shift}
\theta = \arg\max_{l\in[0,L-1]}R_{xy}[l],
\end{equation}
where \(R_{xy}[l] = {\sum_{n=0}^{L-1}{x[n]y[(n+l)\bmod L]}}\). \rev{Assuming zero shift,} $y\sim \mathcal{N}(x,\sigma^2I)$, and  \(R_{xy}[l]=R_{xx}[l]+\sum_{n=0}^{L-1}x[n]\varepsilon_i[(n+l)\bmod{L}]\). \rev{Thus,}
%\(R_{xy_i}[j] \sim\ \mathcal{N}(R_{x,x}[j],\sigma_c^2)\),
\(R_{xy}\sim\mathcal{N}(R_{xx},\Sigma)\), and
\begin{equation}
\Sigma_{i,j}=
%E\left[(\sum_{l=0}^{L-1}x[l]\varepsilon_i[(l+i)\bmod{L}])(\sum_{l=0}^{L-1}x[l]\varepsilon_i[(l+j)\bmod{L}])\right] 
\sigma^2\sum_{l=0}^{L-1}x[l]x[l+j-i\bmod L],
\end{equation}
with diagonal entries \(\Sigma_{i,i}=\sigma^2\|x\|^2 := \sigma_c^2\). \rev{The probability of estimating a shift \(m\) is the probability of the event in which the correlation between \(x\) and \(y\) at entry \(m\) was greater than \revv{all} other entries. In other words,} the PMF of the estimated shift\rev{s} is:
\begin{equation}
p(m) = \text{Prob}\left(R_{xy}[m] > \max_{\substack{n\\n\neq m}}R_{xy}[n]\right).
\end{equation}

\begin{figure}[H]
	% generated with rotation_err_pmf_verification.m
	\includegraphics[width=\linewidth]{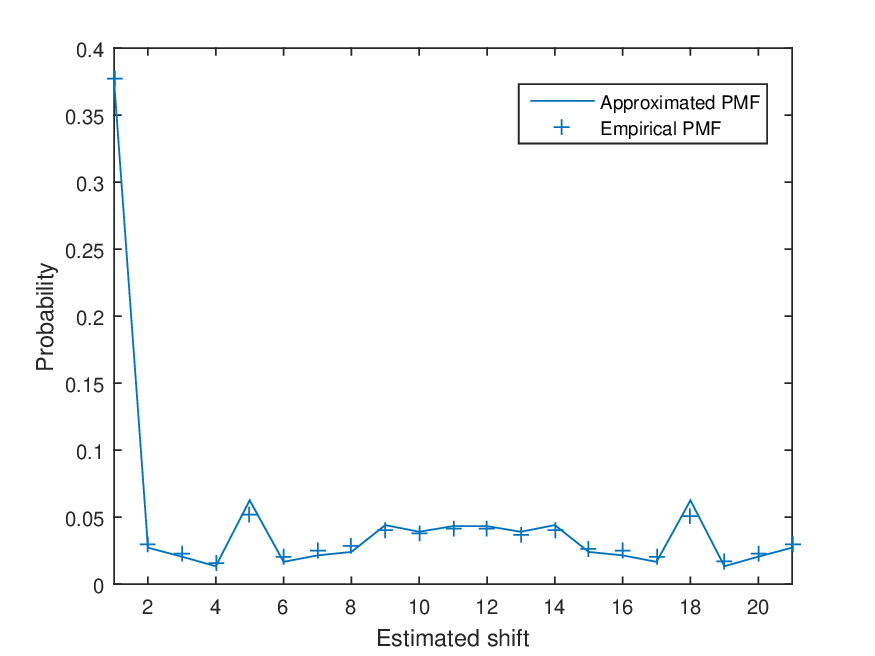}
	\caption{Approximated distribution  \revv{(computed analytically according to (\ref{shift_pmf}))} for a single known realization of \(x\), with \(L=21\) and \(\sigma = 3\), compared against the \rev{empirical} distribution.
%		, for two different signals. In blue, \(x\sim \mathcal{N}(0,I)\), in red \(x\sim \mathcal{N}(0,\Sigma_3)\). The covariance is generated according to a spectral decay model \(\Sigma_b = \Re\{F^*D_bF\}\), where \(D_b=diag\{P^{(b)}\}\), \(P^{(b)}_k=1/(1+k)^b \quad \forall k\in[0,(L-1)/2]\), \(P^{(b)}_k = P^{(b)}_{L-k}  \quad \forall k\in[(L-1)/2,L-1]\). As can be seen in the figure, the approximation deteriorates as the covariance of the signal becomes non diagonal.
	}
	\label{template_matching_pmf}
\end{figure}

In general, \(\Sigma\) is not \revv{a diagonal matrix}.
% since \(\Sigma_{i,j}\neq 0 \quad \forall i\neq j\).
However, since \(x\sim\mathcal{N}(0,I)\), we can approximate \(\Sigma\) (for large L) \revv{by} \(\Sigma \approx \sigma_c^2 I\). Under this approximation, it follows that the elements of \(R_{xy}\) are independent, with \(R_{xy}[j] \sim\ \mathcal{N}(R_{xx}[j],\sigma_c^2)\). 
\rev{
\rev{\revv{Let us denote} \(A_m = R_{xy}[m]\) and \(B_m = \max_{\substack{n\\n\neq m}}A_n\). Thus, the probability density function of \(A_m\) is:}
\begin{equation}\label{folded_pdf}
f_{A_m}(u) = \frac{1}{\sqrt{2\pi\sigma_c^2}}e^{-\frac{(u-R_{xx}[m])^2}{2\sigma_c^2}},
\end{equation}
\revv{and} the cumulative distribution function is
\begin{equation}\label{folded_cdf}
F_{A_m}(u) = \frac{1}{2}+\frac{1}{2}erf\left({\frac{u-R_{xx}[m]}{\sqrt{2\sigma_c^2}}}\right).
\end{equation}
\rev{Since we assumed that the random variables \(A_0,...,A_{L-1}\) are independent, the cumulative distribution function of \(B_m\), their maximum excluding \(A_m\), is given by:}
\begin{equation}\label{max_dist}
\begin{aligned}
F_{B_m}(u) =
P\left(B_m<u\right)
=\prod_{\substack{n=0 \\n \neq m}}^{L-1}{F_{A_n}(u)}.
\end{aligned}
\end{equation}

%Using the joint probability function
%\begin{equation}
%\begin{aligned}
%P\left(A_m > B_m \right) = 
%\int_{-\infty}^{\infty}{\int_{-\infty}^{u}{f_{A_m,B_m}(u,v)dv}du}.
%\end{aligned}
%\end{equation}
Since \(A_m\) and \(B_m\) are independent, \(f_{A_m,B_m}(u,v)=f_{A_m}(u)f_{B_m}(v)\), and \tamirrev{thus}
\begin{equation}
\begin{aligned}
P\left(A_m > B_m\right)=
\int_{-\infty}^{\infty}f_{A_m}(u)\int_{-\infty}^{u}{f_{B_m}(v)dv}du.
\end{aligned}
\end{equation}
Using the cumulative distribution function, \tamirrev{we can also write:}
\begin{equation}
\label{pmf_before}
\begin{aligned}
P\left(A_m > B_m\right)=
\int_{-\infty}^{\infty}{f_{A_m}(u)F_{B_m}(u)du}
\end{aligned}.
\end{equation}
Substituting (\ref{folded_pdf}), (\ref{folded_cdf}) and (\ref{max_dist}) into \rev{(\ref{pmf_before}) yields}:
}
%\begin{equation}\label{shift_pmf}
%p(m)=\int_{0}^{\infty}{
%	\frac{1}{\sqrt{2\pi\sigma_c^2}}\left(e^{-\frac{(x-R_{xx}[m])^2}{2\sigma_c^2}}+e^{-\frac{(x+R_{xx}[m])^2}{2\sigma_c^2}}\right)
%	\prod_{\substack{n=0 \\n \neq m}}^{L-1}{\frac{1}{2}\left(erf\left({\frac{x-R_{xx}[n]}{\sqrt{2\sigma_c^2}}}\right) +
%		erf\left({\frac{x+R_{xx}[n]}{\sqrt{2\sigma_c^2}}}\right)
%		\right)}dx}
%\end{equation}
\begin{equation}
\label{shift_pmf}
\begin{aligned}
p_\theta[m]\approx\int_{-\infty}^{\infty}{
	\frac{1}{\sqrt{2\pi\sigma_c^2}}e^{-\frac{(u-R_{xx}[m])^2}{2\sigma_c^2}}}\times
\\
{
	\prod_{\substack{l=0 \\l \neq m}}^{L-1}{\left(\frac{1}{2}+\frac{1}{2}erf\left({\frac{u-R_{xx}[l]}{\sqrt{2\sigma_c^2}}}\right)\right)
	}du}
\end{aligned}.
\end{equation}
Figure \ref{template_matching_pmf} shows the approximated distribution and the \rev{empirical} distribution for a single realization of \(x\), with \(L=21\) and \(\sigma=3\). The \rev{empirical} distribution was computed using \(100,000\) samples.
%Even in the trivial case of 1-D MRA observations with zero rotations and a known signal \(x\), the pmf had to be approximated for a closed form solution.
%In practice, the signal \(x\) is unknown and distributed according to some \(p(x)\), the problem might be either 1-D or 2-D, and the synchronization method considered is the angular synchronization in which rotation of one observation depends on all observations. In this case, a closed form expression for the pmf of the estimated rotation is beyond reach, therefore learning methods should be adopted.

\begin{figure}[H]
	% generated with rotation_err_pmf_verification.m
	\includegraphics[width=\linewidth]{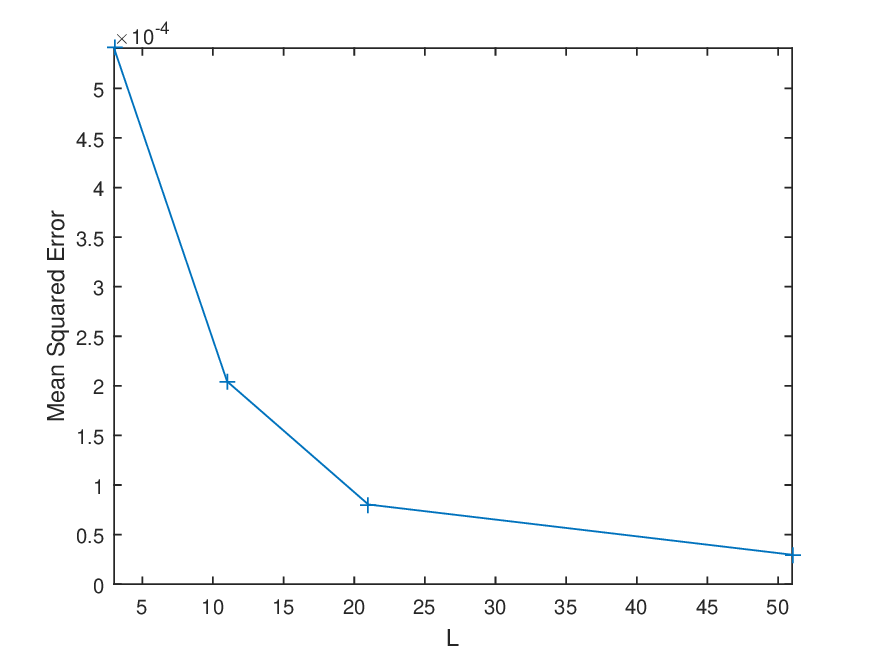}
	\caption{Mean squared error between the approximated distribution  \revv{(computed analytically according to (\ref{shift_pmf}))} and the \rev{empirical} distribution.}
	\label{pmf_approximation_error}
\end{figure}

Figure \ref{pmf_approximation_error} shows the mean squared error between the \rev{empirical distribution} (\revv{computed} over 100,000 samples) and the approximated distribution \revv{(computed analytically according to (\ref{shift_pmf}))}, averaged over 10 different \rev{signal} realizations, as a function of \(L\) for \(\sigma=3\). The \rev{average} is taken across the different shifts and realizations. As can be seen, the approximation gets better as \(L\) increases.

\section{M-Step for MAP-EM with prior on the distribution}
\label{Mstep}
In this section, we \rev{derive} the maximization step  
%\((x_{t+1},\rho_{t+1})=\arg\max_{x,\rho}Q(x,\rho|x_t,\rho_t)\)
\((a_{t+1},\rho_{t+1}) = \arg{\max_{a,\rho}{Q(a,\rho|a_t,\rho_t)}}\), where \(Q\) is given by~(\ref{expectation_step}). 
%From the structure of Q, it follows that optimization over \(\rho\) can be done independently. 
By substituting the prior on the distribution of rotations given in Section \ref{GibbsPrior}, the Lagrangian is given by: 
\begin{equation}
\begin{aligned}
&\mathcal{L}(\rho,\lambda) = \\
&\sum_{l=0}^{L-1}{
	W_t[l]
	\log{\rho[l]}
}
-\gamma \sum_{l=0}^{L-1}{\bar{\rho}[l]\log{\frac{\bar{\rho}[l]}{\rho[l]}}}
-\lambda\left(
\sum_{l=0}^{L-1}{\rho[l]}-1\right).
\end{aligned}
\end{equation}
%Differentiating by \(\rho\) and equating to zero:
\rev{Setting the derivatives to zero \revv{yields}} 
\begin{equation}
\frac{\partial\mathcal{L}(\rho,\lambda)}{\partial\rho[l]} = 
\frac{W_t[l]}{\rho[l]}
+\gamma\frac{\bar{\rho}[l]}{\rho[l]}
-\lambda=0.
\end{equation}
Solving for \(\rho\):
\begin{equation}
\label{RhoLambda}
\rho[l] = \frac{1}{\lambda}(W_t[l]+\gamma \overline{\rho}[l]).
\end{equation}
Differentiating by \(\lambda\) and equating to zero:
\begin{equation}
\label{DiffByLambda}
\frac{\partial\mathcal{L}(\rho,\lambda)}{\partial\lambda}=
\sum_{l=0}^{L-1}{\rho[l]}-1=0,
\end{equation}
which enforces that \(\rho\) is normalized \revv{to one}. By substituting~(\ref{RhoLambda}) into~(\ref{DiffByLambda}), we get \(\lambda=\sum_{l=0}^{L-1}{(W_t[l]+\gamma\bar{\rho}[l])}\), which results in:
\begin{equation}
\rho_{t+1}[l] = \frac{W_t[l]+\gamma\bar{\rho}[l]}
{\sum_{l=0}^{L-1}{(W_t[l]+\gamma\bar{\rho}[l])}}.
\end{equation}
%Maximization over \(x\):
%\begin{equation}
%x_{t+1}=\arg\max_{x}
%\log p(x)-\sum_{j=1}^N{\sum_{l=0}^{L-1}\frac{w_t^{l,j}}{2\sigma^2}\|R_{l}x+\mu-y_j\|_2^2}
%\end{equation}
%Since \(x \sim\ \mathcal{N}(0,1)\), \(\log p(x)=-\frac{1}{2}\|x\|^2\).
%\begin{equation}
%x_{t+1}=\arg\min_{x}
%\sigma^2\|x\|^2+\sum_{j=1}^N{\sum_{l=0}^{L-1}w_t^{l,j}\|R_{l}x+\mu-y_j\|_2^2}
%\end{equation}
%Differentiating by \(x\) and equating to zero results in:
%\begin{equation}
%\sigma^2 x +\sum_{j=1}^N{\sum_{l=0}^{L-1}w_t^{l,j}R_l^T(R_{l}x+\mu-y_j)}=0
%\end{equation}
%Optimizing over \(x\):
%\begin{equation}
%x_{t+1}=\arg\max_{x}
%\log p(x)-\sum_{j=1}^N{\sum_{l=0}^{L-1}\frac{w_t^{l,j}}{2\sigma^2}\|R_{l}x-y_j\|_2^2}
%\end{equation}
%Maximizing over \(a\):
%\begin{equation}
%\label{a_max}
%a_{t+1}=\arg\max_{a}
%\log p(a)-\sum_{j=1}^N{\sum_{l=0}^{L-1}\frac{w_t^{l,j}}{\sigma^2}\|a \circ e^{-\I\frac{2\pi  l}{L}\bar{k}}-v_j\|^2}.
%\end{equation}
%Since \(x \sim\ \mathcal{N}(0,\Sigma)\), \(\log p(x)=-\frac{1}{2}x^T\Sigma^{-1} x\).
\tamirrev{Maximizing over \(a\) with \(a \sim\ \mathcal{CN}(0,\Gamma_a)\) results in:}
%Since \(a \sim\ \mathcal{CN}(0,\Gamma_a)\), \revv{it follows that} \(\log p(a)=-a^H\Gamma_a^{-1} a +\text{constant}\). Substituting into~\rev{(\ref{a_max})}, we get:
%\begin{equation}
%x_{t+1}=\arg\min_{x}
%\sigma^2 x^T\Sigma^{-1} x+\sum_{j=1}^N{\sum_{l=0}^{L-1}w_t^{l,j}\|R_{l}x-y_j\|_2^2}
%\end{equation}
\begin{equation}
a_{t+1}=\arg\min_{a}
\sigma^2 a^H\Gamma_a^{-1}a +\sum_{j=1}^N{\sum_{l=0}^{L-1}w_t^{l,j}\|a \circ e^{-\I\frac{2\pi  l}{L}\bar{k}}-v_j\|^2}.
\end{equation}
%\rev{Setting the derivative to zero}:
%\begin{equation}
%2\sigma^2\Sigma^{-1} x +\sum_{j=1}^N{\sum_{l=0}^{L-1}w_t^{l,j}R_{l}^T(R_{l}x-y_j)}=0
%\end{equation}
%\begin{equation}
%\sigma^2\Gamma_a^{-1}a +\sum_{j=1}^N{\sum_{l=0}^{L-1}w_t^{l,j} e^{\I\frac{2\pi  l}{L}\bar{k}}\circ (a \circ e^{-\I\frac{2\pi  l}{L}\bar{k}}-v_j)}=0
%\end{equation}
\tamirrev{Setting the derivative to zero and solving for \(a\) yields:}
\begin{equation}
a_{t+1} = 
\left(NI + \sigma^2\Gamma_a^{-1}\right)^{-1}\sum_{j=1}^{N}{\sum_{l=0}^{L-1}
	{
		w_{t}^{l,j}e^{\I\frac{2\pi l}{L}\bar{k}}\circ v_j
	}
}.
\end{equation}
%\begin{equation}
%x=\left(NI + 2\sigma^2\Sigma^{-1}\right)^{-1}
%\sum_{j=1}^N{\sum_{l=0}^{L-1}w_t^{l,j}R_{l}^Ty_j}
%\end{equation}

%\section*{Acknowledgment}

\ifCLASSOPTIONcaptionsoff
  \newpage
\fi

\end{document}